\documentclass[12pt]{article}

\usepackage{epsfig,euscript,amsfonts,amsmath,amssymb}

\newcommand{\br}{{\bf {R}}}

\newcommand{\Tr}{\rm{Tr}}

\author{Andrei Khrennikov\\International Center for Mathematical Modeling \\
in Physics, Engineering, Economics, and Cognitive Science\\
Linnaeus University, V\"axj\"o-Kalmar, Sweden }
\title{``Einstein's Dream''  -- Quantum Mechanics as Theory of Classical Random Fields}

\begin{document}

\maketitle

\abstract{This is an introductory chapter of the book in progress on quantum foundations and incompleteness of quantum mechanics. 
Quantum mechanics is represented as statistical mechanics of classical fields.}

\bigskip

{\bf Preface}

\bigskip

This book is dedicated to Einstein's vision of physics and specifically his hope for what quantum theory could and, in his view, should be. In particular, two of Einstein's dreams about the future of quantum theory are realized in this book: a reduction of quantum randomness to classical ensemble randomness and the total elimination of particles from quantum mechanics (QM) – the creation of a field model of quantum phenomena. Thus, contrary to a number of the so-called "no-go" arguments and theorems advanced throughout the history of quantum theory (such as those of von Neumann, Kochen-Specker, and Bell), quantum probabilities and correlations can be described in a classical manner.

There is, however, a crucial proviso. While this book argues that QM can be interpreted as a form of classical statistical mechanics (CSM), 
this classical statistical theory is not that of particles, but of fields. This means that the mathematical formalism of QM must be translated into the mathematical formalism of CSM on the infinite-dimensional phase space. The infinite dimension of the phase space of this translation is a price of classicality. From the mathematical viewpoint this price is very high, because in this case the theories of measure, dynamical systems, and distributions are essentially more complicated than in the case of the finite-dimensional phase space found in a CSM of particles. 
However, at the model level (similar to quantum information theory) one can proceed with the finite-dimensional phase space
by approximating physical prequantum fields by vectors with finite number of coordinates. To simplify the presentation and by taking into account
that usage of infinite-dimensional analysis (in the rigorous mathematical framework) is still not common in the quantum community, we 
present the basic constructions in finite dimensional (so to say $n$-qubits) Hilbert spaces. (Special sections are devoted to generalization
to the case of fields.)

On the other hand, from the physical and philosophical viewpoints, considering QM as a CSM of fields can resolve the basic interpretational problems of QM. For example and in particular, quantum correlations of entangled systems can be reduced to correlations of classical random fields. From this perspective, quantum entanglement is not mysterious at all, since quantum correlations are no longer different from the classical ones. \footnote{In fact, the situation 
is more complicated. Averages and correlations provided by the 
classical field theory are related to continuous signals, but 
the quantum ones are based on statistics of discrete clicks of detectors.
However, it is possible to discretize continuous signals with the aid of
the {\it threshold type detectors} and transform probabilistic quantities for 
continuous signals into statistics of discrete clicks which coincide 
with quantum ones. The main part of this book is devoted to 
representation of quantum averages and correlations with the aid of continuous 
random signals. Already this step is nontrivial both from physical and mathematical viewpoints.
The corresponding discretization model will be presented , see also  \cite{ARXIV} 
for more details. We call this model {\it threshold detection model} (TSD).}

The main difficulty is that the classical situation is very tricky by itself. All quantum correlations contain the irreducible contribution of a background field ("vacuum fluctuations"). Roughly speaking quantum systems are classical random signals that are measured against a sufficiently strong random background. The data of QM is a result of our ignorance concerning the contribution of this random background. Thus quantum probabilities and correlations are not simply classical quantities. They are obtained from classical quantities by means of a renormalization procedure: a subtraction of the contribution of the background field. Accordingly, the reduction of quantum randomness to classical is not totally straightforward. Nevertheless, it is possible. For example, the otherwise mysterious nature of the Heisenberg uncertainty principle can be resolved in the following way. Quantum dispersions are not simply classical dispersions, but the results of the subtraction of the dispersion of vacuum fluctuations. There is nothing mysterious in the fact that renormalized quantities satisfy this type of inequality. This is not so unusual from the viewpoint of the classical probability theory.

Already Max Planck emphasized the role of a random background field, the concept that was widely used in stochastic electrodynamics.
(In his letter to Einstein he pointed out that spontaneous emission  can be easily explained by taking into account the background field.)   This field also plays an essential role in our model, CSM of classical fields, also termed as {\it prequantum classical statistical field theory} (PCSFT). This model is purely that of the field type. A classical random field is associated with each type of quantum particles. We have, for example, the electronic, neutronic, and protonic fields. The photonic field is simply the classical electromagnetic field of low intensity. 

There is also a deep-going analogy between the present approach and the {\it classical theory of random signals.} Quantum measurements can be described as measurements for classical random signals with a noisy background. This analogy between QM and classical signal theory had been explored from the reverse perspective, for example, in using quantum information theory in the theory of classical Gaussian random signals. Indeed, restricting the present discussion to Gaussian signals alone would significantly simplify the presentation of PCSFT. However, the present book attempts to proceed by considering arbitrary random signals as much as possible.

I hope that the book will stimulate research that aims to demystify QM and to create a purely field model of quantum reality, and even to go beyond QM and find classical wave phenomena behind the basic laws of QM, such as Born's rule for probabilities. In PCSFT, prequantum random fields fluctuate on a time scale that is essentially finer than the time scale of quantum measurements. The fundamental question is whether this time scale is physically approachable remains open. In particular, if the prequantum time scale were the Planck scale, the  PCSFT-level would be inapproachable. There would be no hope to monitor prequantum waves and show how the quantum statistics of  clicks of detectors is produced 
through interaction of such waves with the threshold-type (macroscopic) detectors.
However, if the prequantum time scale is essentially coarser than the Planck scale, one might dream of finding experimental 
confirmation of derivability of quantum laws (which are fundamentally probabilistic) from behaviour of prequantum (random) waves.
Either way, however, PCSFT provides an adequate theoretical model of reduction of QM to CSM of fields.
\medskip

The present wave of interest in quantum foundations is caused by the tremendous development of quantum information science and its applications to quantum computing and quantum communication. Nowadays this interest even increases, because {\it it became clear that some of the difficulties encountered in realizations of quantum information processing are not simply technicalities, but instead have roots at the very fundamental level.} To solve such difficult problems, quantum theory has to be reconsidered. In particular, some prejudices must be discarded; first of all  
the prejudice on completeness of QM. 

 \vspace{\baselineskip}
\begin{flushright}
\noindent V\"axj\"o-Moscow-Tokyo
\hfill {\it Andrei Khrennikov}\\
2008-2012\hfill
\end{flushright}

\newpage

\tableofcontents

\newpage

{\bf Introductory Chapter}

\bigskip

\section{Author's views on quantum foundations}

This section is full of my reminiscences of meetings in V\"axj\"o,
philosophical and historical remarks on views of Hertz,
Boltzmann, Planck, Einstein, Bohr, Schr\"odinger, von Neumann, De
Broglie, von Mises, Lamb and Scully, Lande, Bohm, Mackey, Marshall
and Brafford, Boyer, de la Pena and Cetto, Hiley, Emch, Cole, Ballentine,
Vidman, Peres, Ohya, Accardi, Gill, Fuchs, Plotnitsky, Zeilinger,
Aspect, Rauch, Weihs, Volovich, Holevo, Belavkin, Ozawa, De
Muynck, De Baere, Elitzur, Peres, Greenberger, ...  This section
can hopefully be of interest as containing details of history
of quantum foundations; or the reader can jump directly to
Section \ref{QU} which contains a short introduction to my
``beyond quantum model'' -- {\it prequantum classical statistical
field theory} (PCSFT).

\subsection{Debates in V\"axj\"o}

To better understand what this book is about, it may be useful to
know a little bit about the author's views.
 In the quantum foundations community I am well
known as the organizer of the series of conferences which have been held in
small town V\"axj\"o in the South-East part of Sweden. This town
surrounded by woods and lakes is really a good place for contemplations of kinds, in my case, specifically, about
quantum theory, one of the most exciting theories ever
created. The theory is exciting not only because of its tremendous
advances, but also because of its paradoxical claims and conclusions.

The series of V\"axj\"o conferences on foundations of quantum
mechanics (especially probabilistic  foundations) combines two
subseries: {\it Foundations of Probability and Physics}: 2000, 02,
04, 06, 08, 11 \cite{KV01},  \cite{KV03},  \cite{KV05}, \cite{KV07}, \cite{KV09};
and {\it Quantum Theory: Reconsideration of
Foundations}: 2001, 03, 05, 07, 09, 12 \cite{KV02},  \cite{KV04}, \cite{KV06}, \cite{KV08},   \cite{KV10}. A new series {\it
Advances in Quantum Theory} started in 2010. All the conferences
have been notable not only for the original contributions but also for several exciting
debates that took place there. These debates offered
a great diversity of perspectives on foundations of quantum
mechanics (QM) and its future developments: from the orthodox Copenhagen interpretation \index{Copenhagen interpretation}(which
rejects realism and causality), at one end of the spectrum, to
more realistic views, as advocated by Einstein, at the other end. \index{Einstein}

During the last ten years I have been lucky to meet the world leading
experts in quantum foundations and discuss with them the most
intriguing problems. What surprised me (at least at the first
conferences)? It was the huge diversity of opinions and views on
the very fundamental and old problems. My expectation that by
inviting  great quantum gurus I can get clear answers was
naive. The first conference, {\it Bohmian mechanics 2000}, \index{Bohmian mechanics} was the
total fiasco: two leading representatives of Bohmian school,
Shelly Goldstein \index{Goldstein} and Basil Hiley\index{Hiley}, presented two totally different
interpretations of Bohmian mechanics. Finally, they accused each
other in misunderstanding of Bohm's views (both had very close
connections to David Bohm). My students whom I invited to learn
Bohmian mechanics from its creators were really confused. The
only useful information which I extracted from Bohmian
mechanics 2000 was that Bohmian mechanics does not give new
experimental predictions comparing to conventional QM. Thus,
although formally (mathematically) Bohmian mechanics provides a
finer description of micro processes, it is impossible to design
experiments which will distinguish Bohmian mechanics and QM.

 Similar stories have repeated quite a few times with various fundamental
problems. Only in some cases I was lucky to learn something. For
example, I got the answer to the question: ``What is crucial in
quantum computing: superposition or entanglement?'' I learned that
superposition plays a subsidiary role, the crucial is
entanglement; the classical wave computer (e.g., optical) cannot
beat the classical digital computer.  However, yet another
simple question has never been clarified: ``Do pure states provide
better quantum computational resource than mixed?'' Opinions of
quantum computing gurus did not converge to the common point. And
I can mention a series of similar questions, e.g., ``How dangerous
for quantum cryptography is detectors inefficiency?'' People who
spontaneously answered me that the impact of the detectors
inefficiency can be easily taken into account, a few years later
applied for grants to study this ``very important problem''.

I can mention a series of simple V\"axj\"o-questions without
answers, e.g.: ``What is electron? What is the origin of
discreteness of the electric charge? What is the essence of vacuum
fluctuations? Can the mathematical formalism of QM be applied
outside of physics, e.g., in cognitive science?'' \index{cognitive science}

However,  the most exciting spectacle started each time when the
question of {\it interpretations of the wave function}  \index{interpretation of wave function} attracted
the attention. Finally, I understood that the number of different
interpretations is in the best case equal to the number of
participants. If you meet two people who say that they are
advocates of, e.g., the Copenhagen interpretation of QM, ask them
about the details. You will see immediately that their views on what is
the Copenhagen interpretation can differ very much. The same is
true for other interpretations. If two scientists tell that they
are followers of Albert Einstein's {\it ensemble interpretation,} \index{ensemble interpretation}
ask them about the details... At one of the round tables (after two hours of
debates with opinions for and against completeness of QM) we had
decided to vote on this problem. Incompleteness advocates have won,
but only because a few advocates of completeness voted for
incompleteness. The situation is really disappointing: the basic
notion of QM has not yet been properly interpreted (after 100
years of exciting, but not very productive debates).\footnote{``When
I speak with somebody and get to know their interpretation, I
understand immediately it is wrong. The main problem is that I do
not know whether my own interpretation is right.'' (Theo Nieuwenhuizen) \index{Nieuwenhuizen}This is the
standard problem of participants of V\"axj\"o conferences.} I specifically
appreciate the activity of Arcady Plotnitsky, \index{Plotnitsky}
 philosopher studying Bohr's views, see, e.g., \cite{P2}--\cite{P4}. He
 teaches us (participants of V\"axj\"o conferences) \index{V\"axj\"o conferences} a lot. First of
 all we got to know that the Copenhagen interpretation \index{Copenhagen interpretation} cannot be
 rigidly coupled with Bohr's views. On many
 occasions  Niels Bohr emphasized that QM is not about physical processes in microworld,
 but about our measurements \cite{BR}: ``Strictly speaking, the mathematical formalism of quantum mechanics and electrodynamics
merely offers rules of calculation for the deduction of expectations pertaining
to observations obtained under well-defined experimental conditions specified
by classical physical concepts''.  The basic postulate of the Copenhagen
 interpretation of QM  -- ``the wave function describes the state
 of a quantum system'' (i.e., a concrete system, not an ensemble)
 -- cannot be assigned to Bohr. Then we learned (again from
 Plotnitsky)  that Bohr's views have been crucially changed a few times
 during his life. Thus, there can be found many different Bohr's
 interpretations of QM. Bohr was definitely the  father of the {\it operational 
 interpretation} of QM.  As was already pointed out, Bohr emphasized that the formalism
of QM does not provide the intrinsic description of processes in microworld, it describes only results
of measurements.   Bohr also can be considered as
 one of fathers of the so-called {\it information interpretation} \index{information interpretation} of
 QM: the QM-formalism describes information about micro systems
 extracted by means of macroscopic measurement devices. Heisenberg
(and to some extent Schr\"odinger) shared this viewpoint. Nowadays
the information interpretation of QM became very popular, see,
e.g., \cite{KV02}, \cite{Folse}, \cite{P4}. I can mention Anton Zeilinger \index{Zeilinger} \cite{Zeilinger} and Christopher
Fuchs  \index{Fuchs } \cite{Fuchs}--\cite{Fuchs1} among the active promoters of this interpretation; we can also mention Mermin's paper \index{Mermin} \cite{ME1}.

One of the most commemorable debates, on {\it Bell's inequality,} \index{Bell's inequality}
was ignited by Luigi Accardi, \index{Accardi} see e.g. \cite{AC2} \cite{ACT} for his views,
 and Richard Gill, \index{Gill} see e.g. \cite{Gill}. They put 1000 Euro
for (Luigi) and against (Richard) a possibility to simulate
EPR-correlations in purely classical local framework, later the
sum reached 3000 euro; I, Slava Belavkin, and Inge Helland
agreed to be in the jury. (Very soon we realized that it was a
wrong decision.) Although the positions of both parties involved
in this great debate have not changed much after ten
years of discussions, nor the jury was able to make a
well-grounded decision, this debate had an impact in  the quantum
community, see, for example, \cite{KV1}. The main message of this debate was that Bell's
arguments were not so well justified as it was commonly believed.
A part of Bell's critique \cite{B} against von Neumann's no-go theorem can
be redirected against Bell's own theorem.  And it is not as easy
as it was believed to defend the Bell's position.

As an expert in probability, in general, I agree with Luigi
Accardi. (But I am not sure that computer simulation can play the
role of a crucial argument.) This inequality is a general
statistical test to check a possibility to fit the data collected
in a few experiments into a single {\it Kolmogorov probability
space}, \index{Kolmogorov probability
space} see monograph \cite{KHR-context} for the complete analysis
of this problem, see also \cite{KH2}, \cite{KHI}, \cite{KL3},
\cite{LARS1}, \cite{Manko}--\cite{Manko3}, \cite{KHFR},
\cite{KL4}, \cite{KV1}, \cite{KL8}, \cite{ADC2P}.  This is a test
of a possibility to use one special model of probability theory,
the Kolmogorov model \cite{K}, for the collected data. I recall
that already Lobachevsky \index{Lobachevsky} and Gauss \index{Gauss} planned experiments to check
applicability of the Euclidean model \index{Euclidean model} of geometry, see Section
\ref{Euclidean} for more detail. They proposed concrete tests.
Bell's inequality is a similar test for Kolmogorov's probability
model. We recall that, in fact, Bell's inequality was invented
many years ago by Boole \index{Boole} \cite{Boole}, \cite{Boole1}: to check that
statistical data can be described by a single Boolean algebra (see
also I. Pitowsky \index{Pitowsky} \cite{Pitowsky} for a detailed analysis of the
problem). General statistical tests of Kolmogorovness of data \index{Kolmogorovness of data} were
found by Soviet mathematician Vorob'ev \index{Vorob'ev} \cite{VOR}. Where does
non-Kolmorovness \index{non-Kolmorovness} come from? It is a separate problem. Two
possibilities were mentioned by Bell: violation of locality \index{violation of locality} or
(and) violation of realism. \index{violation of realism} But, in principle, non-Kolmogorovness
can come from various sources different from those mentioned by
Bell, see my monograph \cite{KHR-context}. Thus the honest
position would be that violation of Bell's inequality can only
tell us that the Kolmogorov's model did not pass the test. We
cannot derive the definite conclusion on concrete sources of
non-Kolmogorovness.

It is not easy to understand why physicists do not like such a
viewpoint (with one exception: the famous Soviet experimenter in
quantum optics Klyshko \index{Klyshko} \cite{Klyshko1}). Theoretical physics is about
creation of mathematical models of reality. Elaboration and
experimental verification of the statistical test, the Bell-Boole
inequality, which shows the boundary of applications of one of
mathematical models, Kolmogorov's one, is the great  success! If
one does not like to proceed in such a way, then they should take
other possible sources of non-Kolmogorovness not less seriously
than Bell's sources. For example, my graduate student Guillaume
Adenier \index{Adenier} studied the impact of the so-called {\it unfair sampling} \index{unfair sampling}
\cite{AD}--\cite{AP}, impossibility to guarantee reproducibility of
statistical properties of ensembles used in experiments with
incompatible pairs of orientations of polarization beam splitters. \index{polarization beam splitter}

It is clear that unfair sampling implies non-Kolmogorovness and,
hence, violation of Bell's inequality. And from our viewpoint,
unfair sampling is not a less important source of
non-Kolmogrovness than, e.g., nonlocality. In turn there are
various sources of unfair sampling. One of the most well known is
the inefficiency of detectors. I spoke a few times with Alain
Aspect about this problem. He was not even sure that experimenters
should concentrate efforts to close this ``loophole''. For him,
this source of violation of Bell's inequality can not be
considered on the same level of importance as ``Bell's sources''.
Opposite to him, I think that closing of the ``efficiency ofdetectors loophole'' \index{efficiency of
detectors loophole} is not less important than, e.g., locality
loophole; without this it is totally meaningless to try to
restrict sources of non-Kolmogorovness to ``Bell's sources.'' From
the purely probabilistic viewpoint it is even more natural to
expect that non-Kolmorovness is induced by the cutoff of
ensembles. Therefore, during a last few years, I and Adenier
avertised the EPR-Bell experiment with {\it Tungsten-based
Superconducting Transition-Edge Sensors} \index{Superconducting Transition-Edge Sensors}  (W-TESs) -- the
ultra-sensitive microcalorimeters, \index{microcalorimeters} see, e.g., \cite{DEM_KHR} for
an experimental proposal.\footnote{Paul Kwiat \index{Kwiat} tried to design such
an experiment in his lab a few years ago. He promised me a talk on
such an experiment at V\"axj\"o conferences 2006, 07, but, as I
understood, his group did not overcome technical problems. Marco
Genovese works on this problem right now; Anton Zeilinger \index{Zeilinger}
recently, February 2010,  told me that  his group will soon start
such experiments. It is clear that this problem (unfair sampling
for the Bell's tests with photons) cannot be solved with the
threshold type detectors, \index{threshold type detectors} photomultipliers tubes \index{photomultipliers tubes} -- PMTs,
avalanche photodiodes -- APDs,visible light counters -- VLPCs.} An
interesting general test of fair sampling was elaborated in
\cite{ADEXP}. Unfortunately, we were not able to convince
experimenters to do this test (in spite of a number of promises,
it has never been done). I also point to the really important
contribution of Jan-Ake Larsson \index{Larsson} \cite{LARS1a} to the study of the
probabilistic structure of the problem of the detectors
efficiency.

Another source of unfair sampling is the use of the
{\it time window} in the EPR-Bell experiments; it also makes
cutoff of ensembles inducing non-Kolmogorovness, see
\cite{HPL10}--\cite{HPL3}, \cite{Gill}, \cite{LARS2}, \cite{Raedt}, \cite{Raedt1},  \cite{PROJ2}
for different viewpoints on this problem.
Unfair sampling need not be reduced to ensemble cutoff, as due to
inefficiency of detectors or time window. \index{time window}  Unfair sampling can take
place for 100$\%$-efficient detectors and practically zero time
window. It appears naturally in models of the EPR-Bell experiment
taking into account parameters of measurement devices \index{parameters of measurement devices} \cite{KHR-context}, \cite{KHRBELL}.
In such models randomness induced by preparation procedure is combined with randomness induced by measurement device.

\medskip

The majority of participants of V\"axj\"o conferences believed
that QM, as it stands now,  will, for a long time to come or even
indefinitely, remain a correct, or indeed the correct, theory
within its proper scope. On this view, improvements in our
experimental technology would not alter the essential features of
QM in its present form, although new exciting experimental and
theoretical findings, including of foundational nature, are
possible. Indeed, even in this group, there was no consensus not
only on whether there is a single correct interpretation of QM but
also on whether we really have a proper foundation of the theory,
say, of the type we have in relativity theory. Several papers
presented in V\"axj\"o explored this question and possible new
foundational approaches to the standard version of QM.\footnote{I
can mention  V. Belavkin, B. Coecke, W. De Muynck, C. Fuchs, R.
Schack, M. Appleby,   A. Plotnitsky, L. Hardy,  M. D'Ariano, A.
Elitzur, W. Zurek, P. Busch, D. Greenberger, R. Balian,  K.
Svozil, J. Smolin, M. Ozawa, A. Peres, D. Mermin, S. Stenholm, J.
Summhammer, P. Lahti, I. Volovich.}

On the other hand, a number of participants actively promoted
alternative models, which might enable us to {\it go beyond the
standard QM.}  In particular, some of them considered QM as an
emergent theory and envisioned, ``dreamed of,'' the possibility of
a reduction of quantum randomness (viewed as irreducible by the
orthodox interpretation) to classical randomness, for example, of
the type found in classical statistical physics.\footnote{I can
mention G. `t Hoot,    T. Elze, C. Garola, M. Davidson, T. Boyer,
T. Nieuwenhuizen, S. Gudder, G. Emch, B. Hiley, C.  Wetterich, D. Cole, L.
de la Pena, G. Adenier,  H. D. Doebner,  A. F.  Kracklauer, Ch. Roychoudhuri.}  The present author
belongs to this (``minority'') group and considers {\it QM as an
approximation, possibly a very good one, of a more fundamental
theory of microscopic reality.}

I emphasize that the reduction of quantum
randomness to classical randomness, e.g., consideration of quantum systems
as classical random signals (as in the present book), does not imply a kind of
comeback to Laplacian determinism. Consider the standard Brownian motion. Suppose
that we are not able to take into account the effect of the random background for an individual particle. In such a
situation we are not able to predict the trajectory of this particle,
even if the initial conditions are determined with high precision. Nevertheless,
the impossibility to predict the trajectory is not interpreted as the absence of the trajectory.
(In \cite{AL1} it was shown that classical Brownian motion \index{Brownian motion} exhbits even such a ``fundamentally quantum property''
as entanglement.)

More generally, it may be argued that strategically there are two
main ways {\it to go beyond QM.} The first is to stimulate the
development of the conventional QM, especially by designing new
experiments, in a hope that, sooner or later, QM will reach the
limit of its validity, even within its proper experimental scope
(i.e., as a nonrelativistic theory of quantum phenomena),
``peacefully,'' as it were, as the result of its own, internal
development. The second approach is to pursue a critique of the
conventional approach in order to find its weak points, properly
handling which would require an alternative theory.

I support both approaches; and, as I said, I believe, with
Einstein, that the standard QM will ultimately prove to  be an
{\it approximation of a more fundamental theory,} perhaps based on
a prequantum mathematical model of the type described above. At
the moment, however, I do not think that any available model of
this type provides a viable possibility in this regard.
Mathematics offers great opportunities to explore, to ``play
with,'' various pre-quantum models. However, such mathematical
games are not the same as real physical models, which must
rigorously relate their mathematics to experimentally observed
phenomena. The main problem of prequantum approaches, including my
own, is the lack of proposals for realistic experiments that could
demonstrate that the QM-formalism describes micro-reality only by
a way of approximations\footnote{My classical field-type model predicts that even the basic rule of QM,
Born's rule, is an approximate rule.} that can be superseded by a better theory.
Nevertheless, these approaches offer a possible new trajectory for
future development of quantum theory, and the main aim of
V\"axj\"o conferences was to explore such trajectories.

Overall, V\"axj\"o conferences have played an important role in the
ongoing investigation of quantum foundations and possibilities
of new discoveries in QM and beyond it\footnote{See, e.g., \cite{AA}--\cite{ACT1}, \cite{AICH}--\cite{AL}, \cite{APPL},
\cite{AP}, \cite{ATM}, \cite{Bacci}, \cite{BL3}--\cite{BELAV}, \cite{BIAL},
\cite{BUS1}, \cite{CAT}--\cite{DAV}, \cite{Bae3}, \cite{LAP1}--\cite{DEM4}, \cite{Folse}--\cite{Fuchs1},
\cite{Gill}, \cite{Gudder1}, \cite{HARD}, \cite{HEL}--\cite{HIES}, \cite{HUD}, \cite{KLA} \cite{KV0}, \cite{KV1}, \cite{KL55}, \cite{KIM}
\cite{LARS1}, \cite{LARS1},  \cite{LOUB}, \cite{ME1}, \cite{NH}, \cite{PERREZ}, \cite{P2}, \cite{SCH}, \cite{Sta}, \cite{Svozil2}, \cite{Volovich},
\cite{Weihs}.}, for example, in quantum
field theory. Several additional aspects and high points of the
conferences are worth mentioning here. The {\it interaction of a
large number of experimenters} at the conferences (e.g., A. Aspect,
H. Rauch, G. Weihs, M. Genovese, S. Kulik, C. Roychoudhuri, M.
Zukowski, F. De Martini, A. Zeilinger, F. Sciarrino, B. C. Hiesmayr, C. Roos,...) will, hopefully,
facilitate the aim of designing new foundational experiments,
especially but other tests of possible
violations of Bell's inequalities, the approach that dominated
this field for a long time. The discussions and debates on
the foundations of quantum information theory (especially quantum
cryptography and computing) that took place during the conference
will, undoubtedly, contribute to further developments
in this important field of research. Indeed, quantum information
theory can be considered as a great experiment to test the
validity of the main principles of quantum mechanics and its
interpretation. Finally,  V\"axj\"o conferences offered
significant new insights into the question of quantum probability,
which remains essential to quantum foundations, however one
pursues them.

\subsection{The role of probability}

{\bf Main message:} {\it Probability in QM should be taken
seriously; this is a tricky notion; even ``classical probability''
can induce rather counter-intuitive (paradoxical) conclusions.
Therefore a rigorous mathematical presentation of probabilistic
statements of QM is very important.}

\medskip

In fact, {\it quantum probability} \index{quantum probability} or, to be more precise, the
difference between quantum and classical probabilistic models was
the starting point of my interest in quantum foundations.
Graduated from the Department of Mechanics and Mathematics of
Moscow State University I was well trained in probability theory;
on a few occasions I was lucky to meet the founder of the modern
probability theory Andrei Nikolaevich Kolmogorov \index{Kolmogorov} \cite{K}.  In
particular, he communicated to Doklady USSR  paper \cite{SML}
(written under supervision of
 Oleg Georgievich Smolyanov) \index{Smolyanov}  devoted to a generalized
model with complex valued probabilities. My PhD-thesis \cite{KHPD1} was
devoted to the theory of {\it probability on infinite-dimensional
spaces} (including distributions on Hilbert spaces) with
applications to mathematical problems of quantum field theory \cite{KHPD2}--\cite{KHPD5}.

This pathway to quantum theory, namely, through theory of
continual integral, secured me for a long time from terrible
problems and prejudices related to quantum foundations which
everybody meets immediately in standard textbooks. The continual
integral approach, especially its Euclidean version, induced an
illusion that quantum theory, at least quantum field theory, is
about integration on infinite-dimensional spaces. During quite
long time I was completely sure that the infinite number of the
state space dimensions is the main point of departure of QM from classical
mechanics.

The main surprise for me in von Neumann's book \cite{VN} was the notion of {\it irreducible quantum randomness}.
\index{irreducible quantum randomness}
Von Neumann \index{von Neumann } sharply distinguished classical and quantum randomness. The first one, which is exhibited everywhere, besides
the QM (in classical statistical mechanics, economics, finances, biology, engineering), is a consequence
of the impossibility to take into account all parameters describing a system and its interaction with
a measurement device (or environment). Although it is difficult and sometimes even really impossible to specify the values
of these parameters (so to say ``hidden variables''), there are no doubts of their existence. For example,
in statistical mechanics the Liouville equation describes dynamics of
 the probability distribution on a phase space. It is very difficult
to solve the corresponding system of Hamiltonian equations
describing trajectories of millions of individual particles.
Nevertheless, there are no doubts that these particles really move
in physical space. The easiest way to describe diffusion
(including the Brownian motion) is to solve the Fokker-Planck
(and, in general, the direct Kolmogorov) equation for the
probability distribution on configuration space. Still,
it is also possible to solve the corresponding stochastic
differential equation and obtain the description of dynamics
by means of trajectories of individual particles, the classical
stochastic process.

In contrast, it has been claimed that quantum randomness cannot be
reduced to our lack of knowledge of ``hidden variables''. It
became rather fashionable to claim that QM has demonstrated the
violation of laws of classical probability theory, see, e.g.,
Richard Feynman \index{Feynman} \cite{Feynman}, p. 2: ``But far more fundamental was the
discovery that in nature  the laws of combining probabilities are
{\it not}  those of the classical probability theory of Laplace.''

Such a viewpoint supports the illusion
that something mystical goes on in QM, since the probability laws
which has been valid everywhere are violated in the microworld. In my
first studies I clarified this problem\footnote{See  \cite{MI1}--\cite{MI3},  \cite{KHI}, \cite{KHFR}, \cite{KL5},
\cite{KL51}, \cite{KL8}, \cite{KL5}, \cite{KL51}, \cite{KH2}, \cite{KHI}, \cite{KL4}, \cite{KL11}, \cite{KL12}, \cite{KLA}--\cite{KL20},
\cite{KL28}, \cite{KL37}, \cite{KL40}, \cite{KHR-context}.}.  One cannot speak about probability without specifying a
mathematical model of probability. In particular, the notion of
``classical probability'' is related to a variety of mathematical
models: Kolmogorov's measure-theoretic, von Mises' frequency,
subjective probability and so on \cite{MI1}--\cite{MI3}, \cite{KHI}. Quantum probability
is nothing else than a probability described by Dirac-von Neumann
model \cite{Dirac}, \cite{VN}: a complex Hilbert space; wave functions and, more generally,
density matrices as states; self-adjoint operators as observables;
Born's rule. Therefore, when one speaks about matching or
mismatching of classical and quantum probability, the classical
model and, what is very important, {\it matching rules
should be specified}.

At the first stage of my exciting journey into the probabilistic
foundations of QM, I found that there is no contradiction between
the QM probabilistic model and von Mises' frequency probability
theory  \cite{KHI}, \cite{KHFR}, \cite{KL5},
\cite{KL51}, \cite{KL8}, \cite{KL5}. And this is not surprising.  The von Mises \index{von Mises}
approach \cite{MI1}--\cite{MI3},  \cite{KHI} is a very general empiric approach providing the
probabilistic model for statistical data collected in experiments.
In fact, von Mises (in 1919) elaborated the probabilistic model
based on the empiricist ideology: statistics of outcomes of
experiments is described by frequencies of realizations. Thus,
{\it the classical probabilistic model of R. von Mises does not
contradict QM.} This was well known already to von Neumann, who
proposed to consider von Mises theory as the probabilistic ground
of QM.\footnote{``However, the investigation of the physical quantities related to a single object $S$ is not the only thing which can be done  --
especially if doubts exist relative to the simultaneous measurability of several quantities. In such cases it is also
possible to observe great statistical ensembles which consist of many systems $S_1,..., S_N$ (i.e., $N$ models of $S,$
$N$ large). (Such ensembles, called collectives, are in general necessary for establishing probability theory as the
theory of frequencies. They were introduced by R. von Mises, who discovered their meaning for probability theory, and who built up
a complete theory on this foundation.)'' See \cite{VN}, p. 298.}
The approach of von Mises is especially useful to describe statistical data collected 
in quantum experiments. It reflects the temporal structure of an experimental
data-stream. And a typical quantum experiment has the well defined temporal structure.
This is a run of preparations and corresponding detections. 

The main difference between von Mises' and von Neumann' views was
that von Mises did not claim that additional variables responsible for randomness could not be
introduced.

\subsection{To Hilbert space from probability}

{\bf Main message:} {\it The Hilbert space model provides a linear space representation of
probabilistic data.}

\medskip

In his famous book \cite{Mackey} Mackey \index{Mackey} started in purely
empiricist manner, i.e., with probabilities collected in various
(in general incompatible) experiments, and tried to find
conditions inducing representation of data by complex probability
amplitudes, or in the abstract framework by normalized vectors in
the complex Hilbert space.\footnote{I called this problem the
``inverse Born problem'' \cite{KLJJ}, \cite{KHR-context}, \cite{Nyman}.
Born's rule \index{Born's rule} obtains probability
from the wave function. We are interested in production of a
complex amplitude from probabilistic data. This amplitude has to
satisfy to the ``direct Born's rule.''} Unfortunately, he did not
succeed completely; finally, the complex Hilbert space
representation was postulated and this decreased the value of
Mackey's studies. I spent a few years by working to complete
Mackey's program and I developed the general contextual approach
based on families of probabilities obtained by measuring of
various observables for various contexts. I found an algorithm
representing probabilities (under special, but natural conditions)
by complex amplitudes -- {\it quantum-like representation
algorithm} \index{quantum-like representation
algorithm} (QLRA); probabilities and amplitudes are coupled by
Born's rule, i.e., the squared amplitude coincides with the
probability. Thus, the complex Hilbert space was not postulated (as
was done by everybody, from Dirac \cite{Dirac} and von Neumann \cite{VN} to Mackey \cite{Mackey}), but
appeared in the constructive way \cite{KHR-context}.\footnote{Moreover,
it was found that the formalism of QM is too restrictive for a
general empiricist model: not all probabilities are represented by
complex amplitudes. \index{complex amplitudes} Some of them are represented by the so-called
hyperbolic amplitudes \index{hyperbolic amplitudes} \cite{KL13}, \cite{KL19}, \cite{KL31}, \cite{KHR-context}, \cite{KHR-finance}
(taking values of the form $z=x+jy, x, y \in
{\bf R}, j^2=+1)$ or mixed hyper-complex amplitudes. Thus the
complex Hilbert space is too restrictive for the linear
representation of the general empiricist (contextual) model; the
hyper-complex Hilbert space provides the proper base.}

\subsection{Against completeness}

{\bf Main message:} {\it In the past, various models of reality have been claimed to be final (complete). The most
known examples are Euclidean geometry (see, especially Kant \cite{Kant}) and Newtonian mechanics. However, sooner or later
such scientific myths died. QM is the latest myth of a complete theory.}

\medskip

My next action towards demystification of the QM-formalism was
against Bohr's thesis on {\it completeness of QM.} I remark that
von Neumann's statement of irreducibility of quantum randomness is
nothing else than the probabilistic performance of Bohr's
statement of {\it completeness of QM.} \index{completeness} By
Bohr QM provides the finest possible description of micro
phenomena; a finer description of the state of a quantum system
than given by the wave function is totally impossible. To justify
completeness of QM, Bohr need not any ``no-go theorem'' (such as
von Neumann's, Kochen-Specker's or Bell's theorems \cite{VN},
\cite{KH}, \cite{B}). He was completely fine with Heisenberg's
uncertainty relation. \index{von Neumann theorem}
\index{Kochen-Specker theorem} \index{Bell theorem}

Einstein \index{Einstein} had never accepted Bohr's \index{Bohr} thesis on completeness of QM.
All his life he dreamed of creation of a new fundamental theory of
micro phenomena. He was sure that the wave function does not
provide the complete description of the state of an individual
quantum system. Einstein was the father of the {\it ensemble
interpretation} of the wave function as describing statistical
properties of an ensemble of  systems created by some preparation
procedure.\footnote{This interpretation was later elaborated by
Leslie Ballentine \index{Ballentine} \cite{BL}, \cite{BL1} who used the termn {\it statistical
interpretation.} \index{statistical
interpretation} Unfortunately, this terminology is rather
misleading, since it had been used by von Neumann, too: the wave
function, although assigned to the state of an individual system,
expresses statistics of measurements (but this statistics is
coupled to irreducible randomness).} Part of the quantum
community borrowed Einstein's ensemble interpretation of the wave
function and combined it peacefully with Bohr's thesis of
completeness of QM. They support the {\it operational or
empiricist interpretation} of QM \cite{Lud}, \cite{BUS},  \cite{Ozawa}, \cite{Ozawa1}, \cite{Holevo0}, \cite{Holevo},
\cite{Klyshko19},
\cite{DEV}, \cite{DARIANO} and consider QM
formalism as a story on preparation and measurement procedures. \index{preparation procedures}
\index{measurement procedures}
At  this stage the positions of Bohr and Einstein coincide.
However, Einstein did not see any fundamental barrier to complete
QM (to know what goes on behind rough macroscopic preparation
and measurement procedures.  Majority of people
using the operational interpretation \index{operational interpretation} do not support Einstein's
views. In contrast to Einstein, they do not dream of new, more
sensitive preparations and measurements which will show that QM
provides only an approximative description of phenomena. But even
this interpretation is better than the orthodox Copenhagen
interpretation, according to which the wave function provides  the complete
description of the state of a micro-system.

Thus {\it the first dream of Albert Einstein was to  reduce
quantum randomness to classical randomness by creating a finer
description of micro-system's state than given by the wave
function.} It was the dream of creation of a kind of classical
statistical mechanics for microsystems. Einstein did not specify
prequantum, so to say hidden, variables. They need not be the canonical pair of $(q,p),$ position and momentum. Classical
prequantum statistical mechanics has to be based on some sort of
deterministic dynamics. For example, one may expect to find a
micro-analog of the Liouville equation and the underlying Hamilton
equations. We emphasize
that Einstein, whose contribution to theory of Brownian motion is
well known, did not dream of the comeback to the Laplacian
determinism. (An essential part of book \cite{EI} was devoted to
critique of  mechanical determinism.)\index{mechanical determinism} The prequantum deterministic
dynamics is definitely influenced  by classical randomness of
the initial conditions and random background.

\subsection{Einstein's dream of the pure field model}
\label{EPURE}

{\bf Main message:} {\it Particles are illusions and fields are reality.}

\medskip

It is less known that Einsteinian ``beyond quantum world'' \index{beyond quantum}  was
not imagined as a micro-copy of our macroscopic world populated by
particles. Einstein's greatest dream was to {\it eliminate
particles totally from coming fundamental theory.} New theory
should be a purely field model of reality:  \index{field model of reality} no particles, but only
fields (or may be just one field). He considered particles as the
relict of the old mechanistic model of reality. It may be not so
well known, but for Einstein QM was not a theory too novel (so that he
even could not understand it, as some people claim), but, in
contrast, it was too old fashioned to be considered a new
fundamental theory.

In \cite{EI} he discussed a
lot Bohr's principle of complementarity, so to say, wave-particle
duality. He was not happy with the quantum jargon mixing waves and
particles. Einstein was sure that the wave-particle duality will
be finally resolved in favor of a purely wave model:

\medskip

``But the division into matter \index{matter} and field \index{field} is, after the recognition
of the equivalence of mass \index{mass} and energy, \index{energy} something artificial and
not clearly defined. Could we not reject the concept of matter and
build a pure field physics? What impresses our senses as matter is
really a great concentration of energy into a comparatively small
space. We could regard matter as the regions in space where the
field is extremely strong. In this way a new philosophical
background could be created. Its final aim would be the
explanation of all events in nature by structure laws valid always
and everywhere. A thrown stone is, from this point of view, a
changing field, where the states of greatest field intensity
travel through space with the velocity of the stone. There would
be no place, in our new physics, for both field and matter, field
being the only reality. This new view is suggested by the great
achievements of field physics, by our success in expressing the
laws of electricity, magnetism, gravitation in the form of
structure laws, and finally by the equivalence of mass and energy.
Our ultimate problem would be to modify our field laws in such a
way that they would not brake down for regions in which the energy
is enormously concentrated. But we have no so far succeeded in
fulfilling this program convincingly and consistently. The
decision, as to whether it is possible to carry it out, belongs to
the future. At present we must still assume in all our actual
theoretical constructions two realities: field and matter.'', see
the book of Einstein and Infeld  \cite{EI}, p. 242-243. Then they
discussed QM and a possibility to interpret the wave function\index{wave function}, the
probability wave\index{probability wave}, as a physical field\index{physical field}:

``For one elementary particle, electron or photon, we have
probability waves in a three-dimensional continuum, characterizing
the statistical behavior of the system if the experiments are
often repeated. But what about the case of not one but two
interacting particles, for instance, two electrons, electron and
photon, or electron and nucleus? We cannot treat them separately
and describe each of them through a probability wave in three
dimensions, just because of their mutual interaction. Indeed, it
is not very difficult to guess how to describe in quantum physics
a system composed of two interacting particles. We have to descend
one floor, to return for a moment to classical physics. The
position of two material points in space, at any moment, is
characterized by six numbers, three for each of the points. All
possible positions of the two material points form a
six-dimensional continuum and not a three-dimensional one as in
the case of one point. If we now again ascend one floor, to
quantum physics, we shall have probability waves in a
six-dimensional continuum and not in a three-dimensional continuum
as in the case of one particle. Similarly, for three, four, and
more particles the probability waves will be functions in a
continuum of nine, twelve, and more dimensions. This shows clearly
that the probability waves are more abstract than the
electromagnetic and gravitational field existing and spreading in
our three-dimensional space,'' \cite{EI}, p. 290-291.

This discussion is very important for our further studies of a
possibility of creation of purely wave picture of physical
reality. It was directly emphasized that one of the main problem
is the impossibility to realize quantum ``waves of probability''
for composite quantum systems on physical space. Recently this
problem was solved by the author of this book, see \cite{KL559}, \cite{KL560}, \cite{KL564}--\cite{KL566}, \cite{KL568}--\cite{KL571}, and the
solution will be presented in  Chapter 2, Section \ref{CS}. Finally, Einstein and Infeld concluded, \cite{EI}, p. 293:

``But there is also no doubt that quantum physics must still be
based on the two concepts: matter and field. It is, in this sense,
a dualistic theory and does not bring our old problem of reducing
everything to the field concept even one step nearer
realization.''

\medskip

As we have seen, Einstein's picture of electron or  neutron
is very simple: these are fields densely concentrated in small
areas of space. These  are classical fields (not
quantum!). Hence, Einstein was sure that the classical space-time
picture of reality could be combined with QM. (We state again that
classical has the meaning classical field theory and not at all
classical mechanics of particles).

\subsection{Anti-photon}
\label{EPURE1}
{\bf Main message:} {\it Photon is a pulse of classical electromagnetic field.
Discreteness is an illusion produced by detectors.}

\medskip

Now we discuss the notion of photon. It is well known that Albert
Einstein invented the notion of the {\it quantum of light}  which was
later called photon\index{photon}.\footnote{Originally the concept of photon was invented by
physical chemist G. N. Lewis who really considered photons as
light particles that transmit radiation from one atom to another.
Wave-like properties of photon were attributed to guiding ghost
field. See Lamb's ``Anti-photon''  \cite{LAMB}, p. 201-211, for more
details. We underscore the difference, the photon-terminology, unlike the quantum of light terminology,
is not so innocent as one may think. By calling
the quantum of light the photon people emphasized
the role  of a particle picture of
light.}

Bohr was not happy with the invention of light quanta. In
particular, the {\it Bohr-Kramers-Slater theory} \index{Bohr-Kramers-Slater theory}\cite{BR2} was
an attempt to describe the interaction of matter and
electromagnetic radiation without using the notion of photon. We
also mention the strong opposition to the notion of photon from
two fathers of QM: Alfred Lande \index{Lande } \cite{LANDE}, \cite{LANDE1} (in
particular, this name is associated with Lande $g$-factor and the
first explanation for the anomalous Zeeman effect) and Willis E.
Lamb \index{Lamb } \cite{LAMB} (e.g., Lamb shift). Their views on
electromagnetism differ crucially from the view of Albert Einstein
(at least, young Einstein, see Section 1.1.9 for the evolution of
the Einstein views).  The latter wrote in 1910 \cite{EINSTEIN3},
p. 207: ``What  we understand by the theory of ``light quanta''
may be formulated in the following fashion: a radiation of
frequency $\nu$ can be emitted or absorbed only in a well defined
quantum of magnitude $h \nu$. The theoreticians have not yet even
come to an agreement in regard to the following question: Can the
light quanta be accounted for entirely by a characteristic of the
emitting or absorbing substance, or should the electromagnetic
radiation itself be assigned, besides a wave structure, such that
the energy of the radiation itself is already divided in definite
quanta? I believe that I have proven that this latter view should
be adopted.''

Both Lande and Lamb rejected the existence of discrete quanta of
electromagnetic field. They were sure that one can proceed in the so-called {\it semiclassical approach},\index{semiclassical approach}
 describing the interaction of
classical electromagnetic field \index{classical electromagnetic field} with quantum matter, \index{quantum matter} see, e.g., \cite{C1}, \cite{C3}, \cite{C2} and recently
\cite{PhotonK}, \cite{PhotonK1}, \cite{Photon1}, \cite{Photon}.  We cite Lamb
\cite{LAMB}, p. 211: ``It is high time to give up the use of the word
``photon'', and of a bad concept which will shortly be a century
old. Radiation does not consist of particles...''
For adherents of the semi-classical approach quantization of the electromagnetic field is done by detectors; it
is not present in electromagnetic field propagating in the
vacuum.\footnote{The semiclassical approach can describe a number
of quantum effects,  e.g., the photoelectric effect (G. Wentzel
and G. Beck, 1926; see W. E. Lamb and M. O. Scully \index{Scully} \cite{LAMBA} for
more detailed calculations). } We also recall that Max Planck\index{Planck}
opposed Einstein's idea of quantum of light from the beginning, and remained
a champion of the unquantized Maxwell field throughout his life.
In 1907 in a letter to Einstein, he said: ``I am not seeking the
meaning of the quantum of action (light quantum) in the vacuum but
rather in places where emission and absorption occur, and I assume
that what happens in the vacuum is rigorously described by
Maxwell's equations,'' see, for example, \cite{Mukunda}.

We also mention {\it stochastic electrodynamics} \index{stochastic electrodynamics}(SED) --   a
variant of classical electrodynamics which postulates the existence of
a classical Lorentz-invariant radiation field (zero point field,
Marshall and Brafford, Boyer,  de la Pena and Cetto, Coli  see,
e.g., \cite{BOY}, \cite{LAP}--\cite{LAP1}, \cite{DC}, see also \cite{Brida}, \cite{NH}). The presence of this field plays the crucial role
in SED's description of quantum effects. We recall that already in
1911 Planck introduced the hypothesis of the {\it zero point
electromagnetic field} in an effort to avoid Einstein's ideas about
discontinuity in the emission and absorption processes.

It is important for our further considerations, that neither  the
semiclassical approach, nor SED, resolve the wave-particle
dualism.  Neither was it resolved by Bohmian mechanics, the modern
version of De Broglie's double solution approach. Bohmian
mechanics reduces the quantum randomness to classical ensemble
randomness, and particles are the basic objects of this theory.

\subsection{Schr\"odinger's wave mechanics}

{\bf Main message:} {\it Wave mechanics is an alternative to the Laplacian deterministic model of particles' motion}

\medskip

By now, the reader might be wondering that
Schr\"odinger's \index{Schr\"odinger} name has not yet been  mentioned in the discussion
of classical and quantum wave mechanics.  I refrained from it till this chapter to have enough place to consider
not only Schr\"odinger's wave mechanics\index{wave mechanics}, but also
his philosophic doctrine, that played an important role in my
own theory.

At the beginning Schr\"odinger considered the squared wave function
of the electron (multiplied by its electric charge $e)$ as the
density of its charge: $$p(t,x)= -e \vert \psi(t,x) \vert^2.$$ The
solution $\psi(t,x)$ of Schr\"odinger's equation for, e.g.,
hydrogen atom describes oscillations of such electronic cloud,
which induce electromagnetic radiation with frequencies and intensities matching the experiment.\footnote{Unfortunately,
I was not able to find in Schr\"odinger's papers any explanation of
the impossibility to divide this cloud into a few smaller clouds,
i.e., no attempt to explain the fundamental discreteness of the
electric charge.} This picture of a quantum particle as a field,
in this case electronic field, coincides with the picture from
Einstein's field dream. Unfortunately, Schr\"odinger was not able
to proceed in this way. He understood, as well as Einstein, that
already for two electrons the wave function cannot be interpreted
as a field on a physical space: $\psi(t,x,y) (x=(x_1,x_2, x_3), y=(y_1,y_2, y_3)),$ is defined on ${\bf
R}^6.$ Although formally Schr\"odinger gave up and accepted Born's
interpretation of the wave function, he did not like the
Copenhagen interpretation, as Einstein did neither, especially,
Bohr's thesis on completeness of QM. Schr\"odinger dreamed to go beyond
QM, and to refind purely wave resolution of
wave-particle duality. Nor, however, did Schr\"odinger accept Einstein's
ensemble interpretation of the wave function. No wonder why!
Schr\"odinger would rather have a wave associated with an individual
quantum system, and the wave function was the best candidate for
such wave. Therefore, he rejected Einstein's idea to
associate the wave function with an ensemble of quantum systems, see their
correspondence in \cite{AFINE}.

As I mentioned, Einstein did not want the comeback to the
Laplacian determinism, and Schr\"odinger did neither: Schr\"odinger's views on
scientific description of physical reality were based on a well
elaborated approach, the so-called {\it Bild-conception tradition},
see D'Agostino \cite{DAG}, p. 351, for details:

Schr\"odinger called ``the classical ideal of uninterrupted
continuous description'', at both observables' and theoretical
levels, an ``old way'', meaning, of course, that this ideal is no
longer attainable. He acknowledged that this problem was at the
center of the scientific debate in the Nineteenth and Twentieth
centuries as well:

``Very similar declarations...(were) made again and again by
competent physicists a long time ago, all through the Nineteenth
Century and the early days of our century...they were aware that
the desire for having a clear picture necessarily led one to
encumber it with unwarranted details,'' \cite{SHX}, p.24.

I would like now to cite a rather long passage from D'Agostino \cite{DAG}, pp.
351-352, presenting philosophic views of Schr\"odinger on two levels of
description of reality: observational (empiricist) and
theoretical.

\medskip

``The competent physicists are almost certainly Hertz\index{Hertz},
Boltzmann\index{Boltzmann} and their followers. One can thus argue that
Schr\"odinger's two-level conception above is, at bottom and
despite its ``amazing'' appearance, part of the tradition of the
nineteenth-century Bild-conception\index{Bild-conception} of physics, formulated by Hertz
in his 1894 Prinzipien der Mechanik, and also discussed by
Boltzmann, Einstein et altri. He partially inherited this
tradition from his teacher Exner and he deepened this conception
through his intense study of Boltzmann's work. One of the main
features of the above tradition is its strong anti-inductionism.
If theory is not observation-depended - in the sense that it is
not constructed on (or starting from) observations - it
consequently possesses a sort of distinction as regards
observations. This distinction may be pushed to various degrees of
independence. Hertz implied that a term-to-term correspondence
between concepts and observables was not needed when he introduced
hidden quantities among the theory's visible ones. In his often
quoted dictum, Boltzmann asserted that only one half of our
experience is ever experience. At bottom, Schr\"odinger was thus
orthodox in his assertions that theory and observations are not
necessarily related in a term-to-term correspondence and a certain
degree of independence exists between them. However, when he added
the further qualification that a repugnancy might exist between
them, he stretched this independence to its extreme consequences,
introducing a quasidichotomy between a pure theory and an
observational language.

This extreme position was not acceptable to the majority of his
contemporaries and to Einstein in particular. Causal gaps,
even if limited to the observables level, could not be accepted by Einstein
and other scientists. In
fact, Einstein's completeness implied a {\it bi-univocal} correspondence
between concepts and observables. It followed from Einstein's
premises that, if Schr\"odinger's wave function did not correspond
to a complete description of the system, the reason was to be
sought in its statistical (in Einstein's sense!) features: i.e.
Schr\"odinger's wave function refers to an ensemble not to an
individual system. Differently, Schr\"odinger thought that
incompleteness in description was generated by an illegitimate
(due to indistinguishability) individualization of classical or
quasi-classical particles in microphysics. On the other hand,
Schr\"odinger could not accept Heisenberg's and Bohr's
Copenhagenism, because, for him, their position represented a
concession to an old conception of the theory-observations
relation, implying that causality-gaps and discontinuities on the
observation-level would forbid the construction of a complete
theory (a complete model). One can thus argue that Schr\"odinger
considered the fundamental defect of the Copenhagen view its
missing the distinction between the two levels of language, the
descriptive and the purely theoretical level. From the QM
impossibility of a continuous descriptive language on the
observable level, the Copenhagenists would have rushed to conclude
the uselessness of a continuous purely theoretical language.''

\medskip

In this book I present a theoretical (causal and continuous) model
of physical reality, {\it prequantum classical statistical field
theory} -- PCSFT.\footnote{In principle, causality is approachable
in the PCFT-framework, but the situation is quite complicated,
because of the presence of vacuum fluctuations; we shall come back
to this problem in Section \ref{WN}}  Since my starting point was
not the observations,  my model does not rely completely on the
descriptive language of QM, which fact is in total accordance with
views of Boltzmann, Hertz, Exner, and Schr\"odinger on relation
between theoretical and observational models. The correspondence
between concepts of PCSFT and QM is not straightforward, see Chapter \ref{threshold}
for coupling of PCSFT and QM through a measurement theory for PCSFT.

Let us return to the views of Schr\"odinger on QM
and physical reality. I cite from Lockwood \cite{Lockwood}, pp.
385-386:

``Two possibilities then present themselves. One possibility (a)
is that individual physical systems do, after all, possess
determinate states in essentially the classical sense. That is to
say, the classical dynamical variables do have well-defined values
at every moment, arbitrary precise simultaneous knowledge of which
is, however, in principle unattainable. Consequently, we have to
fall back on statistical statements. The assertions of quantum
mechanics should accordingly be understood to refer, as in
statistical mechanics, to the distribution of values of these
variables within an ideal ensemble of similarly prepared systems.
Schr\"odinger assumed this to be Einstein's position. The
other possibility (b) is that the quantum-mechanical description,
as embodied in the $\psi$-function, is a complete specification of
an objectively ``fuzzy'' state. On this conception, quantum
mechanics does offer a model of reality; but the model it presents
us with is of an objectively ``blurred'' reality.\index{blurred reality} The difference
between these two interpretations, Schr\"odinger regards as
analogous to the difference between an out-of-focus photograph of
something with perfectly sharp outlines, and a properly focused
photograph of something lacking sharp outlines, such as a patch of
fog. Having set up these alternatives, Schr\"odinger then,
disconcertingly, proceeds to argue that neither is tenable.''

In fact, the viewpoint to QM generated by PCSFT in combination with the corresponding threshold detection model (TSD) does not match neither with (a), the classical statistical mechanical viewpoint to 
QM, nor with (b), the completeness viewpoint to QM.  The aforementioned two level description of reality based 
on the combination of theoretical and observational models is sufficiently close to the one given by PCSFT/TSD.
However, opposite to Schr\"odinger and other adherents of the Bild concept, I think that all basic features of observational 
model have to be derivable from the theoretical model.  
This is a good place to point  to a general scientific methodology 
which was advertized during many years by Atmanspacher and Primas \cite{ATM}. Any scientific theory is based on two levels of description of reality:  ontic (reality as it is) and epistemic (the image of reality obtained with the aid of a special class of observables). The QM-formalism is an example of an epistemic model. In this framework PCSFT can be considered as the ontic model and TSD as the epistemic model which is equivalent to QM.  However, I am not fine with the notion of the ontic model as a model of objective reality, i.e., reality existing 
independently of our observational abilities. I rather prefer a two level description with two collections of variables, so to say fine and coarse
variables. The coarse-variables  are already approachable and the fine ones are yet not, but they will be in future. 
The coarse variables are determined by the fine ones. However, in PCSFT/TSD approach this realition is very tricky, see Section \ref{objectivity}
for further discussion.

\subsection{Bohr-Kramers-Slater theory}

In paper \cite{BR2} Bohr, Kramers, and Slater\index{Bohr-Kramers-Slater theory} (BKS) tried to treat
the interaction of matter and electromagnetic radiation without
photons. By their model atoms produce a virtual field (induced by
virtual oscillators) which induces the emission and absorbsion
processes. This virtual field contains contributions of all atoms
and hence each transition in a single atom is determined by
processes in all atoms nearby. The BKS-theory can be coupled with
PCSFT. In the latter any ``quantum particle'' is represented by a
classical random field. In particular, any atom is nothing else
than an atomic field. A group of atoms induces a collective atomic
field. Therefore we might try to interpret the virtual BKS-field
as the real atomic field of PCSFT. Any transition in atom (by the
QM-terminology from one level to another) is a completely causal
process of evolution of this field. Fields of various types of
``quantum particles'' can interact with each other or better to
say there is a single fundamental prequantum field which have
various configurations: photonic (electromagnetic), electronic,
atomic,.... In PCSFT we stress a role of the background field,
vacuum fluctuations. This field is present even in the absence of
``quantum particles''. The presence of the background field may
solve one of the main problems of the BKS-theory, namely, possible
violation of the law of conservation of energy on the individual
level: the impossibility to account for  conservation of energy in
a process of de-excitation of an atom followed by excitation of a
neighboring one. In PCSFT the energy of the fundamental prequantum
field is not changed in this process.

The BKS-theory was an attempt to unify wave and particle pictures
on the basis of the classical field theory. This was an attempt of
{\it causal continuous description} of quantum jumps in the
processes of absorbtion and emission. We remark that Bohr
elaborated his principle of complementarity only because he was
not able to construct a satisfactory causal field-type model.
Later he advertised completeness of QM \cite{BR0}, \cite{BR}.
Roughly speaking he tried to stop studies similar to his own in
1924th. (The Freudian background of such behavior is evident.)

\subsection{On the evolution of Einstein's views: from classical electrodynamics to photon and back}

Einstein\index{Einstein} has views, as presented respectively in Sections
\ref{EPURE} and \ref{EPURE1}, appear to be in conflict.  On the
one hand, as discussed in Section \ref{EPURE}, he was the
discoverer of the particle of light, the photon, as it eventually
became known, and thus advocated the particle-like model of the
behavior of light in certain circumstances, a view confirmed by
the Compton scattering experiment in 1923, shortly before the
discovery of quantum mechanics. On the other hand, as discussed in
Section \ref{EPURE1}, he championed the classical-like field model
as the best, if not the only model, for fundamental physics,
which, given the continuous character of the classical field
theory, is difficult to reconcile with the concept of photon. This
discrepancy leads one to suspect that these two positions reflect
the views of two different ``Einsteins,'' especially given that
they correspond to two different periods of Einstein�s work, the
first, roughly between 1905-1920, and second, from roughly 1920 to
his death in 1955. His book with Infeld\index{Infeld}, discussed in
\ref{EPURE1}, was originally published in 1938 and, thus, it might
be added, was written shortly after the EPR argument, which
solidified Einstein's critical assessment of quantum mechanics.
The evolution of Einstein's views is instructive and one might
sketch this evolution roughly as follows.\footnote{The account of
this evolution sketched here is courtesy of Arkady Plotnitsky
[private communication]. See also Pais \cite{Pais} for a
discussion of the development of Einstein's views on fundamental
physics, from his earlier work to his work on general relativity
and beyond; and for Einstein's earlier views, see Don Howard and
John Stachel, \cite{Howard} and also \cite{EINSTEIN3}. For
Einstein�s later views, see especially both of his contributions
to the Schilpp volume \cite{Schilpp}.}

It may be argued that Einstein's primary model for doing
fundamental physics was had always been Maxwell's electrodynamics\index{Maxwell's electrodynamics}
as a field theory, which grounds special relativity, introduced in
1905, the same year he introduced the idea of photon. It also
appears, however, that his thinking at the time was more flexible
as concerns what type of physical theory one should or should not
use. His approach was determined more by the nature of the
experimental phenomena with which he was concerned, or in his own
later words, his attitude was more ``opportunistic''
\cite{Schilpp}, p. 684, rather than guided by a given set of
philosophical preferences, as in his later works. In this respect,
the term ``opportunistic'' may no longer easily apply to his later
thinking, or at least his opportunism was conditioned by his
philosophical inclinations toward a classical-like
field-theoretical approach to fundamental physics. Einstein
appears to have introduced the concept of photon under the
pressure of experimental evidence, such as that reflected in
Planck's law or the law of photoeffect (for which Einstein was
actually awarded his Nobel Prize). He went further than Planck by
proposing that the photon was a real particle (rather than a
mathematical convenience), the idea that took a while, until 1920s
and much additional experimental evidence, most especially, again,
Compton's scattering experiments, to accept. Intriguingly, not
only Planck but also Bohr was among the skeptics, and Bohr only
accepted the idea in view of these experiments. Planck, who, as
discussed earlier, strongly resisted Einstein's introduction of
the concept of the photon, had never reconciled himself to the
idea. Thus, it appears that until roughly 1920, Einstein did not
have a strongly held philosophical position of the type he
developed later on, first, following his work on general
relativity (a classical-like field theory) and, secondly and most
especially, in the wake of quantum mechanics. It is worth noting
in this connection that he initially resisted Minkowsky's concept
of spacetime as insufficiently physical, but eventually came to
appreciate its significance, again, especially in view of its
effectiveness in general relativity. It is true, however, that
theoretical physics at the time, including quantum theory (the
``old'' quantum theory), was still more classically oriented, as
against quantum mechanics in the Heisenbergian approach. In
addition, given that, in some circumstances, light would still
exhibit wave behavior, Einstein also believed at the time (until
even 1916 or so), that a kind of new synthesis of the
particle-like and the wave-like theory of radiation would be
necessary. However, this hope had not materialized in any form
that he found acceptable, and he was especially dissatisfied with
Heisenberg's approach \cite{HEI}, developed into the matrix mechanics by Born
and Jordan, or related schemes, such as Dirac's one \cite{Dirac}. The success
of general relativity as a classical-like field theory was
significantly responsible for strengthening Einstein's
field-theoretical predilections, and shaped his program of the
unified field theory (with a unification of gravity and
electromagnetism as the first task), which he pursued for the rest
of his life. The problems of quantum mechanics and his debate on
the subject with Bohr continued to preoccupy him as well, as
reflected in particular in his persistent thinking concerning the
EPR experiment, on which he commented virtually until his death.
His view of fundamental physics following his work on relativity
was also more mathematically oriented than the earlier one. In
particular, he came to believe that it is a free mathematical
conceptual construction, such as those of Riemann's geometry and
tensor calculus in the case of general relativity and indeed of a
similar classical-like field-theoretical type, that should and, he
even argued, will allow us to come closest to capturing, in a
realist manner, the ultimate character of physical reality. He
expressly juxtaposed this approach to that of the
Copenhagen-G\"ottingen approach in quantum mechanics
\cite{Schilpp}, pp. 83-85. In sum, Einstein had come to be
convinced that a strictly field-like theory unifying the
fundamental forces of nature should be pursued. He saw this kind
of theory as the best and even, to him, the only truly acceptable
program for the ultimate theory of nature, while he believed
quantum mechanics to be a provisional theory, eventually to be
replaced by a field theory of the type he envisioned.

It may be remarked that the idea of particle poses difficulties
for this view,  especially the particle nature of radiation,
initially represented in the idea of photon. This is why Einstein
preferred and saw as more promising (than matrix mechanics)
Schr\"odinger's wave mechanics, or why earlier he liked de
Broglie's approach (which he used in his work on the Bose-Einstein
theory). It is true that the latter does retain the concept of
particle and, as such, represents an attempt at a synthesis of the
wave and the particle pictures, which, as noted above, Einstein
contemplated initially. Later on, however, he did not like Bohmian
mechanics, which pursued a similar line of thinking, although his
negative attitude appears to have been determined by a complex set
of factors. Eventually it became apparent that Schr\"odinger's
formalism could not quite be brought under the umbrella of
Einstein's unified field-theoretical program, a la Maxwell,
although in his later years (in 1940s-1950s) Schr\"odinger return
to his initial ideas concerning wave mechanics. Quantum
electrodynamics and then other quantum field theories appeared
even more difficult to reconcile with this approach. Even general
relativity posed certain significant problems for Einstein's
vision, such as singularities, eventually leading to ideas such as
black holes, although the full measure of these difficulties
became apparent only later on, after Einstein's death.

There is thus quite a bit of irony to this history. While Einstein
was  fundamentally responsible for several theoretical ideas that
eventually led others to quantum mechanics, he had developed grave
doubts about quantum mechanics as a ``useful point of departure
for future development'' \cite{Schilpp}, p. 83. Since, however,
our fundamental physics remained incomplete at the time, Einstein
thought that his vision might ultimately be justified. It might
yet be, since our fundamental physics still remains incomplete,
and in particular, is defined by a manifest conflict between
relativity and quantum mechanics or higher-level quantum theories.
It would be curious to contemplate whether Einstein would have
liked something like the string and brane theories, or any other
currently advanced programs for fundamental physics and cosmology.

\section{Prequantum classical statistical field theory: introduction}
\label{QU}

Now I turn to my model, PCSFT, which is based on the unification of
two Einstein's dreams: to reduce quantum randomness to classical
randomness and to create a purely wave model of physical reality.
I emphasize from the very beginning that the majority of
PCSFT-structures are already present in QM, but in PCSFT they obtain a new
(classical signal) interpretation. Therefore the introduction in PCSFT presented
in this section can be considered as a short dictionary that establishes a
correspondence between terms of QM and PCSFT. However, PCSFT not only reproduces
QM, but provides a possibility to go beyond it. Therefore, advanced
structures of PCSFT do not have counterparts in QM.

\subsection{Classical fields as hidden variables}

{\bf Main message:} {\it Quantum randomness is reducible to randomness of classical fields.}

\medskip

Classical fields are selected as the hidden variables\index{hidden variable}.\footnote{PCSFT is not
a deterministic-type model  with hidden variables. By fixing  a classical ``prequantum''
field we cannot determine the values of observables. These values can be predicted with
probabilities which are determined by the prequantum field\index{prequantum field}, see Chapter \ref{MEASUREMENT}
for a measurement theory in the PCSFT-framework.}
Mathematically, they are functions $\phi: {\bf R}^3 \to {\bf C}$
(or, more generally, $\to {\bf C}^k)$ which are square-integrable,
i.e., elements of the $L_2$-space. The latter condition is
standard in the classical signal theory.

In particular, for electromagnetic field, this is just the condition of the
finiteness of energy
\begin{equation}
\label{SEA5}
\int_{{\bf R}^3} (E^2(x) +B^2(x)) dx = \int_{{\bf R}^3} \vert
\phi(x) \vert^2 < \infty,
\end{equation}
where
\begin{equation}
\label{SEA5R}
\phi(x)=  E(x) + iB(x)
\end{equation}
is the {\it  Riemann-Silbertstein vector} \index{Riemann-Silbertstein vector}(the complex representation of
the electromagnetic field).

Thus, the state space of our prequantum model is $H=L_2({\bf R}^3).$ Formally,
the same space is used in QM, but we couple it with
the classical signal theory. For example, the quantum wave function satisfies
the normalization condition
\begin{equation}
\label{SEA4}
\int_{{\bf R}^3} \vert \phi(x) \vert^2 =1,
\end{equation}
but any vector $\phi$ in $H$ can be selected as a PCSFT-state.  These prequantum
waves evolve in accordance with Schr\"odinger's equation\index{Schr\"odinger equation}; formally, the only
difference is that the initial condition $\phi_0$ is not normalized by 1, see
Section \ref{SER}, equation (\ref{SE2}). Thus, these PCSFT-waves are closely related
to Schr\"odinger's quantum waves. However, opposite to Schr\"odinger and to the orthodox
Copenhagen interpretation, the wave function of the QM-formalism is not a state of
a quantum system. In the complete accordance with Einstein's dream of reducibility
of quantum randomness, wave function is associated with an ensemble. The ensemble, however,
not of quantum systems, but the ensemble of classical fields, or, more precisely,
a {\it classical random field}, random signal. It is appropriate to say that,
although our model supports Einstein's views on the origin of quantum randomness, it
also matches von Neumann's views \cite{VN} on individual quantum randomness.
By using ergodicity, see Section \ref{TIME}, we can switch from ensemble description
to individual signal description and vice versa. We state again that such a possibility
of peaceful combination of Einstein's and von Neumann's views on quantum randomness
is a consequence of the rejection of the corpuscular model in the complete accordance
with the views of ``late Einstein.'' (It seems that at first he wanted to reduce
quantum randomness to randomness of ensembles of particles.)

A random field\index{random field} (at a fixed instant of time) is a function $\phi(x,
\omega),$ where $\omega$ is a random parameter. Thus for each
$\omega_0,$ we obtain the classical field, $x\mapsto \phi(x,
\omega_0).$ Another picture of the random field is the $H$-valued
random variable, each fixed $\omega_0$ determines a vector
$\phi(\omega_0) \in H.$ A random field is given by a probability
distribution on $H.$ For simplicity, we can consider a
finite-dimensional Hilbert space instead of $L_2({\bf R}^3)$ (as
people often do in quantum information theory). In this case,
PCSFT considers $H$-valued random vectors, where $H={\bf C}^n.$
(However we strongly emphasize the role of the physical state
space $H=L_2({\bf R}^3),$ see also \cite{Volovich},  \cite{KV1}.)

This is the ensemble model of the random field. In the rigorous mathematical
framework it is based on the {\it Kolmogorov probability space}\index{Kolmogorov probability space} \cite{K}
$(\Omega, {\cal F}, {\bf P}), $ where $\Omega$ is a set and ${\cal F}$ is the
$\sigma$-algebra of its subsets, ${\bf P}$ is a probability measure on ${\cal F}.$
It is always possible to choose $\Omega=H$ and ${\cal F}$ as the $\sigma$-algebra
of Borel subsets\footnote{This is the minimal system of subsets $H$ containing all
balls in $H$ and closed with respect to countable unions, intersections and
complements of sets. In particular, it contains all open and closed sets. However,
the reader with the background in physics can relax:  in this book we shall never use measure-theoretic
constructions at the mathematical level. It is enough to know (without mathematical details) 
about such notions as measure and integral.}
of $H,$ and probability is a measure on the Hilbert space $H.$  We remark that such measures
are used in classical signal theory as probability distributions of random signals.

In the classical signal theory one can move from the ensemble
description of randomness to the time series description -- under the ergodicity
hypothesis, see Section \ref{TIME}. Random signals are widely used e.g. in radio-physics \cite{CL_SIG1};
these are electromagnetic fields depending of a random parameter; by using the Riemann-Silberstein representation
stationary radio-signals can be represented in the complex form: $\phi(x, \omega)= E(x, \omega) + i B(x, \omega).$

\medskip

{\bf Remark 2.1.} Einstein used to make a point that the wave function $\Psi$ is a label
for an ensemble of identically prepared quantum systems. However, it was far from clear
 which statistical characteristics of an ensemble are encoded in $\Psi.$ Obviously, not all of them, since
Einstein lamented that QM is not complete. Our model, PCSFT,
specifies the statistical characteristics are encrypted in $\Psi,$
these are correlations between components of the field. The
correlations are described by the covariance operator of the
probability distribution of hidden variables of the field-type.
This is an important improvement of the statistical interpretation
of QM. Instead of the Einstein's vague statement (see also
Margenau \cite{Margenau} and Ballentine \cite{BL}--\cite{BL4})
about statistical characteristics of an ensemble, we discovered
the classical statistical variable, the covariance operator, which
was formally used in the QM-formalism under the name ``wave
function''. Finally, we remark that people using the operational
interpretation of QM (e.g., Ludwig,  Davis, D'Ariano,  Holevo,
Busch, Grabowski, Lahti,  Ozawa \cite{Lud}, \cite{BUS},
\cite{Holevo0}, \cite{Holevo}, \cite{Klyshko1 9}, \cite{DEV}, \cite{DEM}
\cite{DARIANO}, \cite{Ozawa}, \cite{Ozawa1}) typically proceed
with the ensemble interpretation, too. In contrast to Einstein,
Margenau, and Ballentine, they are sure that $\Psi$ encodes all
possible statistical characteristics of an ensemble, because they
believe in completeness of QM. At the first sight, PCSFT presents
a strong argument against such a viewpoint (introducing a new
statistical characteristic): the covariance operator does not
determine a probability distribution uniquely. Therefore a random
field contains essentially more information than given by the
covariance operator. However, if a prequantum field is Gaussian,
it is completely determined by its covariance operator. (We shall
consider only random fields with zero average.) Thus, for Gaussian
prequantum fields, views of the adherents to the ``orthodox
ensemble interpretation'' can be easily combined with views of the
adherents to the operational approach to QM. (As we see,
surprisingly many contradictions between different interpretations
of QM can be resolved by PCSFT.)

\subsection{Covariance operator interpretation of wave function}
\label{SPSP}

{\bf Main message:} {\it The wave function is not a field of probabilities or a physical
field. It encodes correlations between degrees of freedom of a prequantum random field.}

\medskip

For simplicity, in this introductory section we consider the case of a single,
i.e., noncomposite, system, e.g., the electron (nonrelativistic, since the present
PCSFT is a nonrelativistic theory\footnote{It seems that there are no problems (neither physical nor mathematical) to
develop a relativistic variant of PCSFT. I plan to do this in future.})
and we neglect for a moment (again for simplicity) fluctuations of vacuum which will
play an important role in our further consideration.

In our model the wave function $\Psi$ of the QM-formalism encodes a class of
prequantum random fields having the same covariance operator (determined by
$\Psi$ and determining a unique Gaussian random field.) We state again that we
consider the case of a noncomposite quantum system; for composite systems, e.g.,
for a pair of photons or electrons, the correspondence between the wave function of QM
and the covariance operator of PCSFT is more complicated, see Chapter 2.

\medskip

In this situation the covariance operator \index{covariance operator}(normalized by dispersion) is given
by the orthogonal projector on the vector $\Psi:$
\begin{equation}
\label{SEA6}
D_\Psi= \Psi \otimes \Psi,
\end{equation}
i.e., $D_\Psi u = \langle u, \Psi\rangle\Psi, \; u \in H.$ Thus, $$
D_\Psi=\vert\Psi\rangle\langle\Psi\vert
$$ in Dirac's notation,
i.e.,
$$
D_\Psi u=\langle u \vert\Psi\rangle \; \vert\Psi\rangle.
$$

We also suppose that all {\it prequantum fields have zero average}
\begin{equation}
\label{SEA6P} E\langle y, \phi \rangle= 0, y \in H,
\end{equation}
where $E$ denotes the classical {\it mathematical expectation}
(average, mean value). By applying a linear functional $y$ to the
random vector $\phi$ we obtain the scalar random variable. In the
$L_2$-case we get a family of scalar random variables:
$$
\omega \mapsto \xi_y(\omega) \equiv \int_{{\bf R}^3} \phi(x,\omega)
\overline{y(x)} dx, y \in L_2.
$$
We recall that the covariance operator $D$ of a random field (with zero average)
$\phi \equiv  \phi(x,\omega)$ is defined by its bilinear form
\begin{equation}
\label{COV} \langle Du, v\rangle = E \langle u, \phi \rangle
\langle \phi, v \rangle, u, v\in H.
\end{equation}

Under the additional assumption that the prequantum random fields
are {\it Gaussian}, the covariance operator uniquely determines
the field. Although this assumption seems to be quite natural both
from the mathematical and physical viewpoints, we should be very
careful. In the case of a single system we try to proceed  as far
as possible without this assumption. However, the
PCSFT-description of composite systems is based on Gaussian random
fields\index{Gaussian random field} (Section \ref{CS}, see also
Section \ref{GUAS} for general discussion of a possible physical
origin of Gaussian probability distributions on the prequantum
level.) Let $H={\bf C}^n$ and $\phi(\omega)= (\phi_1(\omega),
...,\phi_n(\omega)),$ then zero average condition (\ref{SEA6P}) is
reduced to
$$
E\phi_i\equiv \int_\Omega \phi_k(\omega) dP(\omega)=0, k=1,..., n;
$$
the covariance matrix $D= (d_{kl}),$ where
$$
d_{kl}=E\phi_k \bar{\phi}_l \equiv \int_\Omega \phi_k(\omega)
\overline{\phi_l(\omega)} dP(\omega).
$$
We also recall that the dispersion of the random variable $\phi$
is given by
$$
\sigma^2_\phi = E \Vert \phi(\omega) - E \phi(\omega)\Vert^2=
\sum_{k=1}^n  E\vert \phi_k(\omega) - E\phi_k(\omega)\vert^2.
$$
In the case of zero average we simply have
$$
\sigma^2_\phi = E \Vert \phi(\omega) \Vert^2= \sum_{k=1}^n  E\vert
\phi_k(\omega)\vert^2.
$$

Here it is always possible to select $\Omega$ (the set of random parameters) as ${\bf C}^n.$
Then, the above integrals will be over ${\bf C}^n.$ In particular, by selecting
Gaussian, complex-valued, random fields we obtain Gaussian integrals over ${\bf C}^n.$

The case $H=L_2$ is more complicated from the measure-theoretic viewpoint, since this space is
infinite-dimensional. In the case of noncomposite systems (i.e., a single photon or electron)
it is also possible to select $\Omega=H,$ i.e., to integrate with respect to all fields of the
$L_2$-class. For composite systems, the situation is more complicated. Here we cannot proceed
without taking into account the background field, that is of the white noise type. And
the well-known fact is that the probability distribution of white noise cannot be concentrated
on $L_2,$ one has to select $\Omega$ as a space of distributions, i.e., to integrate with
respect to singular fields.

We also remark that the random field $\phi(x,\omega)$ corresponding to a pure
quantum state is not $L_2$-normalized. Its $L_2$-norm
\begin{equation}
\label{SEA3} \Vert \phi \Vert^2 (\omega) \equiv \int_{{\bf R}^3}
\vert \phi(x, \omega) \vert^2 dx
\end{equation}
fluctuates depending on the random parameter $\omega.$ We call the quantity
$$\pi_2(\phi)\equiv \Vert \phi \Vert^2$$
the {\it power of the prequantum field (signal)} $\phi.$ This quantity will play
a crucial role in the measurement theory corresponding to PCSFT, see Chapter \ref{MEASUREMENT}.

We shall distinguish the power of a prequantum field from its energy. The latter
is given by the Hamilton function ${\cal H}(\phi),$ (functional (\ref{BBN0}))
which is also a quadratic functional of the prequantum field. However, in contrast
to the ``pure field dependence'' of  $\pi_2(\phi),$ the Hamilton function
${\cal H}(\phi)$ depends on some parameters (mass, charge, external potential).
The $\Vert \phi \Vert^2 (\omega)$ is the power of the $\omega$-realization of
the random field.

\subsection{Quantum observables from quadratic forms of the prequantum
field}

{\bf Main message:} {\it In spite of all no-go theorems (e.g., the
Kochen-Specker theorem), a natural functional representation of
quantum observables exists.}

\medskip

In PCSFT quantum observables are represented by corresponding
quadratic forms\index{quadratic form} of the prequantum field.\footnote{This is true for a
part of PCSFT reproducing QM, cf. Chapter 2, Section \ref{Beyond}
and Chapter \ref{ALPHA} for the PCSFT ``beyond quantum model''.}
A self-adjoint operator $\widehat{A}$ is considered as the symbolic
representation of the PCSFT-variable
\begin{equation}
\label{SEA2}
\phi \mapsto f_A(\phi) = \langle \widehat{A} \phi, \phi \rangle.
\end{equation}
This is a map from the $L_2$-space of classical prequantum fields
into real numbers, a quadratic form.

We remark that $f_A$ can be considered as a function on the phase
space of classical fields: $f_A\equiv f_A(q,p),$ where
$\phi(x)=q(x)+ip(x),$ i.e., it is possible to move from the
complex representation to the phase space representation and vice
versa, see Chapter 3. A crucial point is that the {\it prequantum
phase space} is infinite-dimensional (and the ``post-quantum phase
space'', i.e., the phase space of ordinary classical mechanics is
finite-dimensional).

The average of this quadratic form with respect to the random field
determined by the wave function $\Psi$ coincides with the corresponding
quantum average:
\begin{equation}
\label{SEA1}
\langle f_A \rangle =\langle \widehat{A} \Psi, \Psi \rangle
\end{equation}
or
$$
\langle f_A  \rangle =\langle \Psi \vert \widehat{A} \vert \Psi \rangle
$$
in Dirac's notation.
Here
$$
\langle f_A \rangle= Ef_A(\phi)= \int_H f_A(\phi) d\mu_\Psi(\phi)
$$
is the classical average and $\mu_\Psi$ is the probability distribution of
the prequantum random field $\phi\equiv \phi_\Psi$ determined by the pure
quantum state $\Psi.$ In the real physical case $H$ is infinite-dimensional;
the classical average is given by the integral over all possible classical fields;
probabilistic weights of the fields are determined, in general, non-uniquely, by the $\Psi.$
Thus, the quantum formula for the average of an observable was demystified:
\begin{equation}
\label{SEA}
\langle \widehat{A} \rangle_\Psi \equiv \langle \widehat{A} \Psi, \Psi\rangle
=\int_H f_A(\phi) d\mu_\Psi(\phi)
\end{equation}
It can be obtained via the classical average procedure.

\subsection{Quantum and prequantum interpretations of Schr\"odinger's equation} \label{SER}

{\bf Main message:} {\it Schr\"odinger's equation with random initial conditions
describes dynamics of the physical random field.\footnote{In the biparticle case Schr\"odinger's equation describes
dynamics of the two-points correlation function for field components, see Section \ref{CORPRED}.
}}

\medskip

Before going to the PCSFT-dynamics, we consider the Schr\"odinger equation\index{Schr\"odinger equation}
in the standard QM-formalism:
\begin{equation}
\label{SE1}
i h \frac{\partial \Psi}{\partial t}(t, x)= \widehat{\cal H} \Psi (t, x),
\end{equation}
\begin{equation}
\label{SE1j} \Psi(t_0, x)=\Psi_0(x),
\end{equation}
where $\widehat{\cal H}$ is Hamiltonian, the energy observable. We recall that
Schr\"odinger tried to interpret $\Psi(t,x)$ as a classical field (e.g., the
electron field; the distribution of electron charge in space). However, he gave
up and, finally, accepted the conventional interpretation, the probabilistic one,
due to Max Born.

\medskip

We recall that a time dependent random field $\phi(t, x, \omega)$ is called
a {\it stochastic process} (with the state space $H=L_2)$. Dynamics of the
prequantum random field is described by the simplest stochastic process which
is given by {\it deterministic dynamics with random initial conditions.}

In PCSFT the Schr\"odinger equation, but with the random initial condition,
describes dynamics of the prequantum random field, i.e., the prequantum
stochastic process can be obtained from the mathematical equation which is
used in QM for dynamics of the wave function:
\begin{equation}
\label{SE2}
 i h \frac{\partial \phi}{\partial t}(t, x,
\omega)= \widehat{\cal H} \phi (t, x, \omega),
\end{equation}
\begin{equation}
\label{SE3}
\phi(t_0, x, \omega)=\phi_0(x, \omega),
\end{equation}
where the initial random field $\phi_0(x, \omega)$ is determined by the
quantum pure state $\Psi_0.$  Standard QM gives the covariance operator of this random field.

Roughly speaking, we combined Schr\"odinger's and Born's interpretations:
the $\Psi$-function of QM is not a physical field, but for each $t$ it
determines a random physical field, i.e., the $H$-valued stochastic process
$\phi(t,x,\omega).$

PCSFT dynamics matches standard QM-dynamics by taking into
account the PCSFT-interpretation of the wave-function, see (\ref{SEA6}).
Denote by $\rho(t)$ the covariance operator of the random field
$\phi(t)\equiv \phi(t,x,\omega),$ the solution of (\ref{SE2}), (\ref{SE3}).
Then $$\rho(t)\equiv \rho_{\Psi(t)} = \Psi(t)\otimes \Psi(t),$$
where $\Psi(t)$ is a solution of (\ref{SE1}), (\ref{SE1j}).

Such  simple description can be used only for a single system and
in the absence of fluctuations of vacuum. In the general case of a
composite system, e.g., a biphoton system, in the presence of
vacuum fluctuations Schr\"odinger dynamics of the $\Psi$-function
encodes only dynamics of the covariance operator of the prequantum
stochastic process, see Chapter 2, Section \ref{CS}. The situation is
essentially more complicated than in the case of a single system. We
found that it is possible to construct a few different prequantum dynamics
which match (on the level of correlations) QM-dynamics, see
Section \ref{DYNSCH}.

\subsection{Towards prequantum determinism?}

{\bf Main message:} {\it The background field is
everywhere.}

\medskip

From the PCSFT-viewpoint, the source of quantum randomness is the randomness
of initial conditions (if one neglects vacuum fluctuations\index{vacuum fluctuations}), i.e., impossibility
to prepare a non-random initial prequantum field $\phi_0(x).$

We expect that in future very stable and precise preparation procedures
will be created. The output of such a procedure will be a deterministic field
$\phi(x),$ i.e., random fluctuations will be eliminated.

However, this dream of creating supersensitive ``subquantum''
technologies which would recover determinism on the microlevel may
never come true. In such a case PCSFT will play the role of
classical statistical mechanics of prequantum fields\footnote{In
ordinary classical statistical mechanics the existence of
Hamiltonian dynamics has merely a theoretical value. In real
applications we operate with probability distributions on the
phase space. Corresponding dynamics is described by Liouville
equation.}. Unfortunately, there are a few signs that it really
might happen. First of all, it might be that the scale of
prequantum fluctuations is very fine, e.g., the {\it Planck
scale.} In this case it would be really impossible to prepare a
deterministic prequantum field. And there is another reason. The
PCSFT-model presented up to now has been elaborated for
noncomposite quantum systems, e.g., a single electron. The
extension of PCSFT to composite systems, e.g., a pair of entangled
photons or electrons, is based on a more complicated model of
prequantum randomness, see Chapter 2, Section \ref{CS}. We should
complete the present model by considering  fluctuations of the
background field (zero point field, vacuum fluctuations), in the
same way as in SED. In reality, these are always present.
Therefore Einstein's dream of determinism cannot be peacefully
combined with the presence of the background field. If this field
is irreducible (as a fundamental feature of space), then
deterministic prequantum fields will never be created.\footnote{As I understood from
conversations with Gerard `t Hooft\index{`t Hooft}, in his model \cite{TH}--\cite{TH2} a background random field
which is considered as fluctuations of space-time by itself also plays an important role.
Nevertheless, he claims that at the subquantum level determinism can be completely restored. 
How?} However,
if this background field is simply noise\footnote{Hence the
completely empty physical space can be really prepared,
``distilled from noise''.} which can be eliminated, then we can
dream of the creation of deterministic prequantum fields.
However, a possibility to prepare such fields does not imply 
deterministic reduction of QM. As was already pointed out, 
the inter-relation between prequantum fields and quantum observables
given by  TSD (measurement theory of classical waves with threshold detectors) 
is really tricky, see Section \ref{detection}.

\subsection{Random fields corresponding to mixed states}

{\bf Main message:} {\it A density matrix is the normalized covariance operator of a prequatum random field.}

\medskip

We now consider the general quantum state given by a density operator\index{density operator} $\rho.$
(We still work with noncomposite quantum systems.) According to PCSFT, $\rho$
determines the covariance operator of the corresponding prequantum field (under
normalization by its dispersion)
\begin{equation}
\label{L1}
 D_\rho =\rho.
\end{equation}
Dynamics of the corresponding prequantum field $\phi(t,x,\omega)$ is also
described by the Schr\"odinger equation, see (\ref{SE2}), (\ref{SE3}), with the
random initial condition $\phi_0(x,\omega).$ The initial random field has the
probability distribution $\mu_{\rho_0}$ having zero mean value and the
covariance operator
$$
D(t_0)=\rho_0.
$$
Under the assumption that all prequantum random fields are Gaussian, the initial
probability distribution is determined in the unique way. In the general (non-Gaussian)
case we lose the solid ground. The $\phi_0(x,\omega)$ can be selected in various ways, i.e.,
it can be any distribution having the covariance $D(t_0).$ We could not exclude such a
possibility. It would simply mean that macroscopic preparation procedures are not able
to control even the probability distribution (only its covariance operator).

Denote by $\rho(t)$ the covariance operator of the random field
$\phi(t)\equiv \phi(t,x, \omega)$ given by (\ref{SE2}), (\ref{SE3}) with $\phi_0$
having the covariance operator $\rho(t_0)=\rho_0.$ Then $\rho(t)$ satisfies the
von Neumann equation. However, $\rho(t)$ has the classical probability interpretation
as the covariance operator $D(t).$ In the Gaussian case $D(t)$ determines completely
the prequantum probability distribution.

\subsection{Background field}
\label{WN}

{\bf Main message:} {\it QM is a formalism of measurement with calibrated detectors
(filtering vacuum fluctuations).}

\medskip

In the general PCSFT-framework the randomness of the initial conditions has
to be completed by taking into account fluctuations of vacuum (to obtain a
consistent PCSFT which works both for one particle system and biparticle system).
In our model the background field\index{background field} (vacuum fluctuations\index{vacuum fluctuations}) is of the white noise
type. It is a Gaussian random field with zero average and the covariance operator
$$D_{\rm{background}}= \varepsilon I, \; \varepsilon > 0.$$
It is a stationary field, so its distribution does not change with time.

Consider (by using the QM-language) a quantum system in the mixed state $\rho_0.$
It determines the prequantum random field $\phi_0 \equiv \phi_0 (x,\omega)$ with
the covariance operator $$\tilde{D}(t_0)=\rho_0 +\epsilon I.$$ The value of
$\varepsilon >0$ is not determined by PCSFT, but it could not be too small for a
purely mathematical reason, see Chapter \ref{CHCOMP}, Section \ref{TTHHJJ}.
(Hence QM is a theory of filtration of strong noise.) Now consider the
solution $\phi(t)$ of the Schr\"odinger equation (\ref{SE2}), (\ref{SE3}) with the
initial condition $\phi_0.$ Its covariance operator can be easily found:
$$\tilde{D}(t)=D (t) + \varepsilon I,$$ where $D(t)$ is the covariance operator
of the process in the absence of the background field, $D (t)=\rho(t).$ Here $\rho (t)$
satisfies the QM-equation for evolution of the density operator, i.e, the von
Neumann equation. Thus on the level of dynamics of the covariance operator the
contribution of the background field is very simple: an additive shift. However, on
the level of the field dynamics the presence of vacuum fluctuations changes the
field behavior crucially.

Consider the prequantum random field $\phi_{0}(x,\omega)$ corresponding to a pure
quantum state $\Psi_0.$ Now (in the presence of the background field) the prequantum
random field $\phi_{0}(x,\omega)$ is not concentrated on a one-dimensional subspace
\footnote{In the absence of vacuum fluctuations the covariance operator of the random
field $\phi_\Psi(x, \omega)$ corresponding to a pure state $\Psi$ is given by the
orthogonal projector on $\Psi,$ see (\ref{SEA6}); the corresponding Gaussian measure
is concentrated on a one-dimensional subspace generated by $\Psi.$ Of course, the latter
is valid only for Gaussian prequantum fields.}
$$H_{\Psi_0}= \{ \phi=c \Psi_0: c\in {\bf C} \};$$
the vacuum fluctuations smash it over $H.$

In the canonical QM the background field of the white noise type is neglected; in fact,
it is eliminated ``by hand'' in the process of detector calibration. And it is the right
strategy for a formalism describing measurements on the random background. However, in
an ontic model, i.e., a model of reality as it is, this background field should be taken
into account. Neglecting it induces a rather mystical picture of quantum randomness.

We shall see that in the PCSFT-formalism the background field
plays the fundamental role in the derivation of Heisenberg's
uncertainty relation, see Section \ref{Heisenberg}. Roughly
speaking, Heisenberg's uncertainty is a consequence of vacuum
fluctuations. \footnote{A similar viewpoint on Heisenberg's
uncertainty relation can be found in Hofmann's PhD thesis
\cite{Hofmann} (1999).}

\subsection{Coupling between Schr\"odinger and Hamilton equations}
\label{SER1}

{\bf Main message:} {\it The Schr\"odinger equation is a complex form of the Hamilton equation
for a special class of quadratic Hamilton functions on an infinite-dimensional phase space.}

\medskip

We remark that the Schr\"odinger equation can be written as the system of Hamilton
equations on the (infinite-dimensional) phase space $Q \times P,$ where $Q=P$ is the
real Hilbert space and $H=Q\oplus iP$ is the corresponding complex Hilbert space. The
prequantum field $\phi(x)=q(x) + ip (x),$ where $q(x)$ and $p(x)$ are real-valued
fields (or more generally, they take values in ${\bf R}^m$). Consider the Hamilton
function\index{Hamilton
function}
\begin{equation}
\label{BBN0}
{\cal H} (q,p )=\frac{1}{2} \langle  \widehat{{\cal H}}
\phi,\phi\rangle,
\end{equation}
or, in Dirac's notation,
$$
{\cal H}(\phi)= \frac{1}{2} \langle\phi\vert\widehat{H}\vert\phi\rangle.
$$
see Chapter 3 for details; in PCSFT ${\cal H} (q,p)$ is the energy of the prequantum
field $\phi(x )=q(x) + ip(x).$ The Schr\"odinger equation (\ref{SE2}) can be written as
the system
\begin{equation}
\label{BBN}
\dot{q}=\frac{\partial {\cal H}}{\partial p}, \; \dot{p}=-\frac{\partial {\cal H}}{\partial q},
\end{equation}
see Strochi\index{Strochi} \cite{Strocchi}. From the PCSFT viewpoint, there is no
reason (at least mathematical) to use only quadratic Hamiltonian
functions.  By considering non-quadratic Hamilton functions we
obtain Hamilton systems\index{Hamilton system} connected with the nonlinear Schr\"odinger
equation, cf. \cite{Bialynicki-Birula0}, \cite{Weinberg},
\cite{Gisin0}, \cite{Gisin}, \cite{Doebner}, \cite{Doebner1}.
PCSFT naturally induces a  nonlinear extension of QM, see Sections
\ref{mumu} and \ref{APPDPD}.

\subsection{Nonquadratic functionals of the prequantum field and violation of Born's rule}

{\bf Main message:} {\it Nonlinear, of order higher than two, contribution of the
prequantum field induces violation of Born's rule.}

\medskip

In principle, there is no reason to restrict PCSFT-variables to quadratic
functionals of the prequantum fields, see (\ref{SEA1}). Let us consider an
arbitrary smooth functional\index{smooth functional} $f(\phi), \phi\in H,$ which maps the field
$\phi\equiv 0$ into zero, $f(0)=0.$ Let us also consider a random field
$\phi=\phi(x,\omega)$ corresponding to a quantum density operator $\rho.$
We can find the classical average
\begin{equation}
\label{BBN1}
\langle f \rangle_\mu = \int_H f(\phi) d \mu (\phi),
\end{equation}
where $\mu$ is the probability distribution of the random field. We shall show,
see Chapter \ref{ALPHA}, that this classical average can be approximated by the
quantum average
\begin{equation}
\label{BBN2}
\langle \widehat{A} \rangle_\rho=\rm{Tr} \rho \widehat{A},
\end{equation}
where
\begin{equation}
\label{BBN3}
\widehat{A}= f^{\prime\prime}(0)/2
\end{equation}
is the second derivative of the field functional $f(\phi)$ at the point $\phi=0$
(divided by the factor 2 which arises from the Taylor expansion). If a Hilbert state
space is finite-dimensional, then this is the usual
second derivative. Its matrix ({\it Hessian}\index{Hessian}) is symmetric. If a Hilbert space is
infinite-dimensional (of the $L_2$-type), then the derivatives are so-called variations.
In the rigorous mathematical framework they are Frechet derivatives, that are used,
e.g., in optimization theory. In the latter case the second (variation) derivative is
given by a self-adjoint operator. This is the PCSFT-origin of the representation of
quantum observables by self-adjoint operators.

\medskip

{\it Quantum observables are represented by self-adjoint operators, since they
correspond to Hessians of smooth functionals of the prequantum field.}

\medskip

Thus the QM-formalism gives approximations of classical averages with respect to
the prequantum random fields by approximating field-functionals $f(\phi)$ by the
quadratic terms of their Taylor expansions.

If the functional $f(\phi)$ is linear, $f(\phi)= \langle \phi, y\rangle, y \in H,$
then its QM-image, the second derivative, is equal to zero. Linear field effects
are too weak and they are completely ignored by QM. However, such functionals and
their correlations are well described by PCSFT. Observation of such effects can be
the first step beyond QM, see Section \ref{CORPRED}.

\subsection{Wave comeback -- a solution too cheap?}

{\bf Main message:} {\it Physical space exists!\footnote{By this statement Igor
Volovich has started his talks at V\"axj\"o conferences for ten years criticizing
quantum information theory which practically ignores this fact.} Hence waves
propagating in this space are basic entities of nature.}

It is well known that Einstein was not happy with Bohmian mechanics\index{Bohmian mechanics}. He considered
this solution of the problem of completion of QM as cheap. Recently Anton Zeilinger\index{Zeilinger}
mentioned (in his lecture at the V\"axj\"o conference-2010, ``Advances in Quantum theory'')
that QM may be not the last theory of micro processes and in future a new fundamental
theory may be elaborated. And looking me in the face, he added that those
who nowadays criticize QM and dream of a prequantum theory will be terrified by this
coming new theory, by its complexity and extraordinarity. They will recall the old
QM-formalism, i.e., the present one, with great pleasure, since it was so close to
classical mechanics. A similar viewpoint on a coming prequantum theory was presented
by Claudio Garola\index{Garola} during our dialog on possible ways to proceed beyond QM \cite{Garola}.

PCSFT is a comeback to classical field theory; roughly speaking,
in the spirit of early Schr\"odinger and late Einstein: the
Maxwell classical field theory is extended to ``matter waves''. Of
course, this comeback is not the dream of the majority of those
who nowadays are not afraid to speculate on prequantum models and
criticize the Copenhagen QM. Nevertheless, I do not think that
PCSFT is a cheap completion of the standard QM. I hope that, in
contrast to Bohmian mechanics, Einstein might accept PCSFT as one
of the possible ways beyond QM. In any event the Laplacian
mechanistic determinism was totally excluded from PCSFT; reality
became blurred in the sense of Schr\"odinger \cite{SCHR0},
\cite{SH1}. This is reality of fields and not particles, but still
reality.

\section{Where is discreteness? Devil in detectors?}
\label{dis}

Prequantum variables $f_A (\phi) = \langle \widehat A \phi, \phi \rangle$ have continuous
ranges of values. On the other hand, in QM some observables have discrete spectra. Thus,
although PCSFT matches precisely probabilistic predictions of QM \footnote{In fact,
the situation is more complicated. By considering quadratic prequantum variables $f_A (\phi)$
we obtain the coincidence of prequantum classical and quantum averages, see Chapters 2,3.
However, by considering nonquadratic functionals of prequantum fields we find that the quantum
probability given by Born's rule is just the main contribution to the prequantum (classical)
average.}, it violates the {\it spectral postulate} of QM. How can one obtain discrete spectra?

The continuous field model supports the viewpoint that ``ontic reality", i.e., reality as
it is, is continuous.\footnote{In fact, my viewpoint on a proper mathematical model of
reality is more complicated. Of course, the usage of continuous space-time based on real
numbers is just a way to unify  a huge hierarchy of scales of space and
time. In this book we do not criticize this model, cf. with, e.g., $p$-adic models,
Vladimirov, Volovich, Witten, Freund,  Dragovich, Aref'eva, Frampton, Parisi, Khrennikov,
Zelenov, Kozyrev, see, e.g., \cite{VL}, \cite{KH1}, \cite{KH2}. At the moment we ``just'' criticize Bohr's postulate,
the existence of the fundamental quant of action given by the Planck constant. We predict splitting of
values of quantum observables at  finer (`` prequantum'') time scales.}
{\it Discreteness of some observable data is created by our macroscopic devices}
which split a prequantum signal in a number of discrete channels. Take a polarization beam
splitter (PBS). Consider first a classical signal. Suppose that PBS is oriented at an
angle $\theta$. Then the classical signal is split into two channels. We can label these
channels as ``polarization up", $S_\theta = +1,$ and ``polarization down", $S_\theta = -1,$
(for $\theta$-direction). The only problem is that the classical signal is present in both
channels. Thus we cannot assign to a classical signal (even to a short pulse) a concrete
value of $S_\theta.$ On the contrary, for a ``quantum signal'' (photon),
detectors never click in both channels; we get either $S_\theta = +1$ or $S_\theta = -1.$
This is a standard example of quantum discreteness.

The first comment of this common description is that the situation ``no double clicks''
is never occurred in real experiments, see e.g., \cite{AS}, \cite{Grangier}, \cite{Grangier1},
\cite{AD}, \cite{ADC2P}. There are always {\it double clicks!}
And they are many! They are partially discarded by using the {\it time window.}
However, this is just a remark on the standard measurement procedure. The main point is that
it is possible to produce discrete clicks even from a classical continuous signal by using
threshold-type detectors. My PhD-student Guillaume Adenier performed numerical simulation for
the threshold detection model of classical signals. He reproduced quantum probabilities of
detection and even in a more complicated framework of classical bi-signals interacting with
two PBSs oriented at angles $\theta_1$ and $\theta_2$ the EPR-Bohm correlations;
Bell's inequality was violated, see \cite{AD1}.

In particular, according to our model, electromagnetic field is quantized only in the
process of interaction with matter. This viewpoint matches well views of Lamb
\cite{LAMB}, Lande \cite{LANDE}, \cite{LANDE1}, Kracklauer\index{Kracklauer} \cite{PhotonK},
\cite{PhotonK1}, Roychoudhuri\index{Roychoudhuri} \cite{Photon1}, \cite{Photon}, Adenier \cite{AD1},
people working in SED, e.g. Marshall and Brafford,
Boyer, de la Pena, Ceto, Coli, ...  \cite{BOY}, \cite{LAP}--\cite{LAP1}, \cite{DC}.
However, PCSFT differs
essentially from a rather popular idea that the electromagnetic field is continuous, but
matter is quantized. This viewpoint was stressed  in the books of Lande \cite{LANDE}, \cite{LANDE1}).
PCSFT does not quantize even the matter, the latter also consists of continuous fields
fluctuating on very fine space-time scales. These scales are not yet approachable. In
future we expect to get a possibility to monitor these fields and not only their
averaged images given by quantum particles. SED-like people do not expect this. It
seems that only Albert Einstein might be happy with PCSFT.

\section{On experiments to tests the Euclidean model}
\label{Euclidean}

One of the most famous stories about Gauss\index{Gauss} depicts him measuring the angles of the
great triangle formed by the mountain peaks of Hohenhagen, Inselberg, and Brocken
for evidence that the geometry of space is non-Euclidean. He tested the inequality:
\begin{equation}  \label{GH}
\alpha_{12} + \alpha_{23} + \alpha_{13} < 2\pi,
\end{equation}
where $\alpha_{ij}$ is the angle between the corresponding sides of the triangle.
Gauss understood how the intrinsic curvature of the Earth's surface would
theoretically result in slight discrepancies when fitting the smaller triangles
inside the larger triangles, although in practice this effect is negligible, because
the Earth's curvature is so slight relative to even the largest triangles that can
be visually measured on the surface. Still, Gauss computed the magnitude of this
effect for the large test triangles because, as he wrote, ``the honor of science
demands that one understand the nature of this inequality clearly''.

On the other hand, if the curvature of space was actually great enough to be
observed in optical triangles of this size, then presumably Gauss would have noticed
it, so we may still credit him with having performed an empirical observation of
geometry, but in this sense every person who ever lived has made such observations.
The first person to publicly propose an actual test of the geometry of space was
apparently Lobachevsky\index{Lobachevsky}, who suggested that one might investigate a stellar triangle
for an experimental resolution of the question. The stellar triangle he proposed
was the star Sirius and two different positions of the Earth at 6-month intervals.
This was used by Lobachevsky as an example to show how we could place limits on
the deviation from the flatness of actual space based on the fact that, in a
hyperbolic space of constant curvature, there is a limit to how small a star's
parallax can be, even for the most distant star. The first definite measurement of
the parallax for a fixed star was performed by Friedrich Bessel (a close friend of
Gauss') in 1838, on the star 61 Cygni. Shortly thereafter he measured Sirius (and
discovered its binary nature). Lobachevsky's first paper on the new geometry was
presented as a lecture in Kasan in 1826 followed by publications in 1829, 1835, 1840,
and 1855 (a year before his death). He presented his lower bound for the characteristic
length of a hyperbolic space in the later editions based on the still fairly recent
experimental results of stellar parallax measurements.

The inequality
\begin{equation}  \label{GH1}
\alpha_{12} + \alpha_{23} + \alpha_{13} = 2\pi
\end{equation}
is a geometric analog of Bell's inequality. Violation of (\ref{GH1}), e.g.,
in the form of (\ref{GH}) implies impossibility to use the Euclidean model. In the same
way violation of Bell's inequality implies impossibility to use the Kolmogorov model.

\end{document}